\documentclass[12pt,a4paper,fleqn]{article}

\usepackage[DIV12]{typearea}
\usepackage{amsmath}
\usepackage{amssymb}
\usepackage{amsfonts}
\usepackage{amscd}
\usepackage{graphicx}
\usepackage[footnotesize]{caption2}
\usepackage{cite}
\usepackage[x11names]{xcolor}
\usepackage{mathrsfs}
\usepackage{bbm}
\usepackage{psfrag}

\newcommand{\eVdist}{\kern-0.06667em}

\def\beqra{\begin{eqnarray}}
\def\eeqra{\end{eqnarray}}

\def\beq{\begin{equation}}
\def\eeq{\end{equation}}

\def\beq{\begin{equation}}
\def\eeq{\end{equation}}
\def\bea{\begin{eqnarray}}
\def\eea{\end{eqnarray}}

\renewcommand{\[}{\left[}

\def\Msun{M_{\odot}{\ }}

\allowdisplaybreaks[1]

\begin{document}

\begin{titlepage}
\renewcommand{\thefootnote}{\alph{footnote}}

\begin{flushright}
SISSA 34/2009/EP
\end{flushright}

\vspace*{1.0cm}

\renewcommand{\thefootnote}{\fnsymbol{footnote}}

\begin{center}
{\LARGE\bf A novel determination of the local dark matter density} 

\vspace*{1cm}
\renewcommand{\thefootnote}{\alph{footnote}}

\textbf{
Riccardo Catena\footnote[1]{Email: \texttt{catena@sissa.it}}
\\
Piero Ullio\footnote[2]{Email: \texttt{ullio@sissa.it}}
}
\\[5mm]
{\it
$^{a,b}$SISSA, Scuola Internazionale Superiore di Studi Avanzati,\\
Via Beirut 2-4, I-34014 Trieste, Italy and\\
INFN, Istituto Nazionale di Fisica Nucleare,\\
Sezione di Trieste, I-34014 Trieste, Italy
}

\end{center}

\vspace*{1cm}              

\begin{abstract}
\noindent 
We present a novel study on the problem of constructing mass models for the Milky
Way, concentrating on features regarding the dark matter halo component. We have
considered
a variegated sample of dynamical observables for the Galaxy, including several results
which have
appeared recently, and studied a 7- or 8-dimensional parameter space - defining the Galaxy
model - by implementing a Bayesian approach to the parameter estimation based on a Markov Chain
Monte Carlo method. The main result of this analysis is a novel determination of
the local
dark matter halo density which, assuming spherical symmetry and either an Einasto or an
NFW density profile is found to be around 0.39~GeV~cm$^{-3}$ with a 1-$\sigma$ error bar
of about 7\%; more precisely we find a  $\rho_{DM}(R_0) = 0.385 \pm 0.027\,\rm GeV
\,cm^{-3}$
for the Einasto profile and  $\rho_{DM}(R_0) = 0.389 \pm 0.025\,\rm GeV \,cm^{-3}$ for the
NFW. This is in contrast to the standard assumption that $\rho_{DM}(R_0)$ is about
0.3~GeV~cm$^{-3}$ with an uncertainty of a factor of 2 to 3.
A very precise determination of the local halo density is very important for interpreting
direct dark matter detection experiments. Indeed the results we produced, together with the
recent
accurate determination of the local circular velocity, should be very useful to
considerably
narrow astrophysical uncertainties on direct dark matter detection.

\end{abstract}
\end{titlepage}
\newpage

\section{Introduction}

There is by now incontrovertible evidence that dark matter is the building block 
of all structures in the Universe. This statement is motivated by the successes of the currently
favored  model for structure formation, the $\Lambda$CDM model, in accounting for a variety
of complementary cosmological observations, as well as by dynamical observations performed 
on the largest structures we see in the Universe, namely galaxy clusters, down to low surface 
brightness and dwarf galaxies. Within the picture one obtains when merging such wealth of 
information, dark matter behaves like a dissipation-less and collision-less fluid and it is cold,
i.e. the free-streaming scale associated to dark matter is negligibly small (more precisely, there
are rather tight constraints on the violation of each of these properties). It follows that none of 
the particles in the Standard Model of Particle Physics is suitable to play the role of dark 
matter. Although there are great expectation from the LHC to shed light on physics beyond
the Standard Model and an eventual embedding of a dark matter candidate within it, 
a clean handle on the dark matter puzzle will come only from direct or indirect detection 
signal of dark matter particles within dark matter halos.

There are classes of dark matter candidates for which it is indeed feasible to extract a direct
detection signal, namely a signal of the interaction of a dark matter particle from the local 
population of the Milky Way dark matter halo within a laboratory detector: the most extensively
studied models are those of Weakly Interacting Massive Particles (WIMPs) and axions.
In both cases the signals scale linearly with the local number density of dark matter particles
(and depend also on their local velocity distribution). It is often quoted that such density is 
merely known within a factor of 2 or 3, an uncertainty which limits significantly 
the possibility of cross-correlating, e.g., for a given supersymmetric model, the production cross
section at the LHC with the signal induced by the scattering rate of the lightest supersymmetric 
particle (dark matter candidate) on a target nucleus in an underground detector.
The problem of the not very accurate determination of the local halo density goes back to 
seminal papers in the field in the eighties and nineties, such as the analysis by Gates, Gyuk \& 
Turner~\cite{Gates:1995dw,Gates:1995js}, which are however based on compilations of 
dynamical constraints for the Galaxy that are by now outdated. In this respect, there has been 
significant improvement in the last few years, and it is reasonable to expect that this would have 
an impact on the inferred local halo density.

There are very strong indications that spiral galaxies are embedded into massive dark halos. 
These include the flatness of rotation curves up to their largest measurable radii (see, 
e.g.,~\cite{RTF}), the dynamics of the satellites (see, e.g.,~\cite{LT}), weak lensing 
results (see, e.g.,~\cite{BBS}),  and warping and flaring of the gas layers (see, 
e.g.,~\cite{SC88,M93}). It is not, however, straightforward to derive detailed models for dark 
halos from such dynamical observations. For external galaxies, one options is make a 
compilation of data for different objects, extrapolate trends to be correlated to physical 
properties of the object and constrain the mean functional form for the dark matter profile,
see, e.g., the recent results in~\cite{Portinari:2009ap,Salucci:2009yp}. The focus here is
however on a single object, the Milky Way, and on addressing whether, starting from 
a rather generic model for the Galaxy, a careful cross correlation of all currently available 
data is sufficiently powerful to provide a relevant constraint on quantities like the 
local halo density.

Several authors have considered the problem of building mass models for the Galaxy 
(an incomplete list of references includes ~\cite{BS80,CO81,Gates:1995js,RK88,Dehnen:1996fa,
OM2000,Klypin,widrow}). The standard approach is to describe the Galaxy as a superposition of 
density profile components, and to discriminate among viable ansatz  by comparing with
available dynamical constraints. We follow this same route here, introducing  
a 7- or 8-dimensional parameter space to model the stellar bulge and bar, the stellar and gas 
discs and the dark matter halo. This parameter space is too large to be efficiently surveyed 
with standard scanning procedure, and in fact previous analysis concentrated on few sample
choices of values for the different parameters. We overcome here this difficulty implementing
instead a Bayesian approach to the parameter estimation (see also \cite{widrow}), analogously to what is commonly 
done to estimate cosmological parameters, and also to some recent studies of 
supersymmetric models in a dark matter contest, see, e.g.,~\cite{Martinez:2009jh,Trotta:2008bp}.

The paper is structured as follows: In Section~\ref{data} we introduce the updated set of 
constraints which can be applied to the Galaxy to study the mass decomposition. The mass
model itself is described in Section~\ref{GM}, while Section~\ref{Likelihood} discusses the
definition of the likelihood function we use in our analysis. Results are discussed in 
Section~\ref{res}, concentrating in particular on the local dark matter halo density. 
Section~\ref{concl} concludes.

\section{Dynamical constraints for the Milky Way}
\label{data}

Our data analysis is based on a Bayesian probabilistic inference (see \cite{trotta2} for a recent review on the subject). In a Bayesian approach to the parameter estimation, all the experimental informations are encoded in the likelihood function. In this section we introduce the data used in the construction of our likelihood function. The target of the analysis is the dark matter halo; datasets will be mainly constraining such component and are less focussed on details in the  stellar or gas components. 

\subsection{Terminal velocities}
We start by introducing observables which are related to the Galactic rotation curve. There are 
different methods providing information on its inner or outer portion. Circular velocities at galactocentric distances smaller than the Sun's  can be obtained measuring the motion of gas clouds in the galactic plane and, under the assumption of strictly circular orbits, finding along each line of sight the extreme radial velocity, i.e. the so-called ``terminal velocity''. The 21-cm line of atomic hydrogen and the CO line have been used numerous times in the literature to determine terminal velocities. 
We take advantage of the modeling by Malhotra~\cite{mal} of CO data from Ref.~\cite{KSW} and 
HI data from Refs.~\cite{WW,kerr}.
We restrain to results obtained at $|\sin l| > 0.35$, with $l$ being the longitude in Galactic coordinates, 
since the method assumes circular motion in an underlying axisymmetric potential well, and the  
approximation of axisymmetry is clearly not valid in the region of the Galactic bar, i.e. at radii smaller than, say, 3~kpc: the sample we will consider is then made of 50 data points at negative longitudes and of 61 at positive longitudes.
Moreover, in fits of such inner rotation curve, we will not try to model small scale fluctuations, such as those most probably associated to the spiral arm structure. Following Ref.~\cite{Dehnen:1996fa},
in order  to allow for random and systematic non-circular velocity, we assume a constant error on
terminal velocities of 7~km~s$^{-1}$. 

\subsection{Local standard of rest velocities}
It is not possible to use the terminal velocity method to infer the outer rotation curve, which must be 
derived instead from a sample of tracers for which both distances and velocities are measured. 
We will consider the compilation of 18 regions of high-mass star formation, within inner, local and 
outer spiral arms, for which parallax and proper motion are measured with high precision at the 
VLBI, as provided in the recent work of Ref.~\cite{reid1}; we compare against the compilation 
of velocities with respect to the local standard of rest velocities, or the heliocentric velocities, versus 
the parallax inferred distances, having gauged out the estimated average value for proper 
motion components. Eleven objects out of eighteen are in the outer part of the Galaxy and will be included in the
analysis; the rest is in the inner part and does not add information with respect to terminal velocities.
For each object, the fit will be against the given values of the local standard of rest velocity $v_{\rm lsr}^{i}$, its
proper motions $(\mu_{l}^{i},\mu_{b}^{i})$ and distance $d^{i}$.

\subsection{Velocity dispersions in a tracer population}

A measurement of the rotation curve at large radii (up to about 55~kpc) has been recently 
provided in the analysis of Xue {\it et al.}\cite{SDSS6}. As kinematic tracers, they 
selected $\sim 2400$ Blue Horizontal-Branch (BHB) 
halo stars from the SDSS DR-6 with distances accurate to $\sim 10 \%$. Ten points in the rotation curve at radii from $7.5$ kpc and $55.0$ kpc have been derived by means of a non trivial matching of the observed radial distribution of radial velocities and the analogous distribution obtained from a numerical simulation of the Galaxy. We cannot follow this route; alternatively one can use the BHB dataset to trace the potential of the Galaxy at large radii $\Phi(r)$ (spherical symmetry is assumed in all components here) using Jeans Equation for the second moment of the velocity:
\beq
\frac{d\left(\rho_\star \,\sigma_r^2 \right)}{dr} + 2 \frac{\beta}{r} \rho_\star \,\sigma_r^2 = - \rho_\star \frac{d\Phi}{dr}
\label{jeans}
\eeq 
where $\rho_\star(r)$ is the density profile of halo stars, $\sigma_r$ their radial velocity dispersion and
the anisotropy parameter $\beta \equiv 1 - \sigma_{t}^2/\sigma_{r}^2$, with $\sigma_t$ the 
tangential velocity dispersion. In case of constant non-zero $\beta$, the solution of Eq.~\ref{jeans}
becomes:
\beq
\sigma_r^2(r) = \frac{1}{r^{2\beta}\,\rho_\star(r)} \int_r^{\infty}d\tilde{r}\;\tilde{r}^{2\beta} \rho_\star(\tilde{r}) \frac{d\Phi}{d\tilde{r}} = \frac{1}{r^{2\beta}\,\rho_\star(r)} \int_r^{\infty}d\tilde{r}\;\tilde{r}^{2\beta-1} \rho_\star(\tilde{r}) \Theta^2(\tilde{r})
\label{jeans2}
\eeq 
with $\Theta$ the circular velocity. Xue {\it et al.} suggest to use $\rho_\star \propto r^{-3.5}$ as determined for Milky Way's halo  stars at 10-60~kpc from SDSS data~\cite{bell}; it is also argued,
based on their numerical simulations and for their particular survey volume, that: 
i) the line-of-sight velocity dispersion (the quantity which is measured) 
$\sigma_{l.o.s.} (r) \approx \sigma_r(r)$; and ii) $\beta=0.37$. We will assume that the first  condition holds, while relax the second condition, assuming $\beta$ as a free parameter.

\subsection{Local circular velocity and Galactocentric distance}

The Oort's constants, A and B, naturally appear in the expressions 
for the mean radial velocity and proper motion of astronomical 
objects in circular motion around the galactic center. 
The measure of their difference and sum constrain, respectively, the
local normalization and slope of the rotation curve, {\it i.e.}:
\beq
A-B = \frac{\Theta_0}{R_0} \,; \qquad \qquad A+B = -\left( \frac{\partial \Theta}{\partial R} \right)_0 \,,
\eeq
were $\Theta_0$ is the local circular velocity and $R_0$ is the galactocentric distance of the Sun.

Constraints on the quantity $(A-B)$ have been derived by measuring the proper motions
of different tracers. From the Cepheid proper motions measured by the Hipparcos satellite~\cite{Hipparcos}
it was found $A-B = (27.2 \pm 0.9) \, {\rm km} \, {\rm s}^{-1}\, {\rm kpc}^{-1}$.
Stars of type O, B5 and B6 also provide good samples to determine Oort's constant 
because of their low velocity dispersion \cite{OB_1}.
A combined analysis of the Hipparcos proper motions and ground-based radial velocity
of O-B5 stars gave as a result \cite{OB_1} $A-B = (30.06 \pm 0.98) \, {\rm km}\, {\rm s}^{-1} \,{\rm kpc}^{-1}$.
A similar analysis carried out with O-B6 stars provided the value 
$A-B = (32.0 \pm 2.24) \,{\rm km} \, {\rm s}^{-1} \, {\rm kpc}^{-1}$ \cite{OB_2}
\footnote{We combine in quadrature the errors on A and B reported in \cite{OB_2}.}.
A value of $(A-B)$ at the high end side is favored also by the recent fit of the proper motions of 
a sample of masers in massive star forming regions (one of the sample quoted above for local
standard of rest velocities) to a linear parameterization of the galactic rotation curve, which finds
$A-B = (30.3 \pm 0.9)\, {\rm km} \,{\rm s}^{-1} \,{\rm kpc}^{-1}$ \cite{reid1}. 
 
The ratio $\Theta_0/R_0$ can also be measured independently from Oort's constants, monitoring
the proper motion of the central radio source Sgr $A^*$. 
Assuming that Sgr $A^*$ is  motionless and located
at the Galactic center, its apparent motion is then interpreted as a consequence of
the motion of the Sun with respect to the Galactic center. By removing the solar motion 
in longitude from the reflex of the motion of Sgr $A^*$ in longitude, the authors of \cite{reid2} find:
\beq
\Theta_0/R_0= A-B = 29.45 \pm 0.15 \, {\rm km} \,{\rm s}^{-1} \,{\rm kpc}^{-1}.  
\label{AmB}
\eeq
We take this precise estimate as our reference value for the combination $A-B$.

The quantity $(A+B)$ is less precisely determined. In the literature
not even the sign of the slope of the rotation curve 
[{\it i.e.} -(A+B)] at $R \ge R_0$ seems to be clearly understood. Positive values 
are quoted in \cite{reid1, BB, honma}, negative values in \cite{SDSS6, sofue, OB_1}
and values compatible with zero in \cite{AB0,OB_2}.
$(A+B)$ has been recently estimated through the
kinematics of old M type stars of the thin disk studied with SDSS data \cite{SDSS7}.
The authors find $\frac{\partial \log \Theta_0}{\partial \log R} = -\frac{(A+B)}{(A-B)} = -0.006 \pm  0.016$
\footnote{This number is extracted from a data sample analysis which assumed
Oort's constant estimated by \cite{Hipparcos}.}
which combined with Eq.~(\ref{AmB}) gives 
\beq
A+B = 0.18 \pm 0.47  \, {\rm km} \,{\rm s}^{-1} \,{\rm kpc}^{-1}.  
\label{ApB}
\eeq
We assume Eq.~(\ref{ApB}) as our reference value for the combination $A+B$.

The monitoring of stellar orbits around the central black hole has recently led to breaking the
degeneracy between the black hole mass and value of the distance of the Sun from the Galactic 
center (setting the conversion from measured angular sizes to proper sizes),
obtaining \cite{R0}:  
\beq
R_0 = 8.33\pm 0.35 \,{\rm kpc}.  
\label{r0}
\eeq
Combining this value with Eq.~(\ref{AmB}), one finds $\Theta_0 = (245 \pm 10.4) {\rm km} \, {\rm s}^{-1}$,
i.e.  the present observations seem to favor values of $\Theta_0$ larger then the IAU recommended 
reference value $\Theta_0 = 220 \, {\rm km} \, {\rm s}^{-1}$ \cite{220}. 
Remind that, neglecting non circular motions, $\Theta_0$ is the velocity, measured from the 
Galactic center, of the local system of rest, which is defined as the frame comoving with an observer 
moving along a closed orbit passing through $R_0$. 
Therefore, $\Theta_0$ has to be extracted from the motion of the Sun
with respect to some reference object, {\it e.g.} ${\rm Sgr A}^*$, or a specific stellar population.
We now know that the larger is the velocity dispersion of the reference sample of stars, the smaller is
the local circular velocity estimated from it. This effect goes under the name of asymmetric drift.
In principle, the correct value of $\Theta_0$ should be therefore estimated in the limit of zero 
velocity dispersion of the reference sample of stars.
For these reasons, it has been pointed out in \cite{OB_1} that the 
recommended value  $\Theta_0 = 220 \, {\rm km} \, {\rm s}^{-1}$ is probably underestimated 
because extracted averaging on populations of stars with different velocity dispersion.

\subsection{Local surface mass density}
The total amount of matter in the solar neighborhood can be inferred by the motion of stars in the direction perpendicular to the Galactic plane. Kuijken \& Gilmore~\cite{KG91} have examined local star velocity fields and derived a constraint for the total mean surface density within $1.1~\rm{kpc}$:
\begin{equation}
\Sigma_{|z|<1.1 \rm{kpc}} = (71 \pm 6) \Msun {\rm pc}^{-2}\;.
\label{sigmatot}
\end{equation}
The local surface density corresponding to the visible components only has been estimated instead with star counts~\cite{KG89}:
\begin{equation}
\Sigma_{*} = (48 \pm 8) \Msun {\rm pc}^{-2}\;.
\label{sigmaba}
\end{equation}

\subsection{The total mass at large radii}
The total mass within a sphere of radius $R\gg R_0$ can be estimated from the velocity distribution of the Milky Way's satellites  \cite{Dehnen:1996fa}. We use measurements of the total mass within 50 kpc and 100 kpc to constrain our Galactic model at large radii. The values which we considered in our analysis are \cite{Sakamoto} 
\begin{equation}
M(<50~\rm kpc) = (5.4 \pm 0.25 ) \times 10^{11}~\Msun \;,
\end{equation}
and \cite{Dehnen:1996fa}
\begin{equation}
M(<100~\rm kpc) = (7.5 \pm 2.5)\times 10^{11}~\Msun  \;.
\end{equation}

\section{A mass  model for the Milky Way}
\label{GM}

We present in the following our reference mass model for the Galaxy. The decomposition is 
performed describing each component as either axisymmetric or spherically symmetric.
Depending on the assumptions, the model is fully specified by, in total, 7 or 8 parameters; 
these are listed in table~\ref{priors}.

\subsection{The stellar disc} 

The mass density of the stellar disk is sketched by a one-component thin disc defined by the 
following function:
\beq
\rho_d(R,z) = \frac{\Sigma_{d}}{2 z_{d}} \, e^{-\frac{R}{R_d}} \, \text{sech}^2\left( \frac{z}{z_d}\right)
\;\;\;\; {\rm{with}} \;\;\;\; R<R_{dm}\;,
\label{disk}
\eeq 
where $\Sigma_{d}$ is the disc surface density and $R_d$ ($z_d$) sets a scale of length in the R (z)
direction. This form is in fair agreement with the one suggested by Freudenreich~\cite{Freud} and fitted against COBE photometric maps of the Galaxy. Freudenreich postulated the presence of a hole in the
inner part; such holes seem to be predicted by N-body simulations as a consequence of
the formation of a stable bar. As none of our model discriminators is very sensitive to fine details in 
the distribution of matter towards the center of the galaxy, to simplify calculations of the rotation curves,
we neglect such feature and keep in mind that, in this way, we probably overestimate the density of stars associated to the disc in the inner Galaxy. Moreover, Eq.~(\ref{disk}) describes a flat disc at all radii and
can not take into account deviations from the flat reference plane~\cite{Freud}.
 
Four parameters are introduced in Eq.~(\ref{disk}), however only two will be let vary freely.  We 
fix the vertical height scale to the best fit value suggested in~\cite{Freud},  $z_{d}$~=~0.340~kpc, 
since dynamical constraints we implement are insensitive to a slight variation around this value; 
we also assume that the truncation radius scales weakly with the value of the local galactocentric distance $R_0$,  according to the prescription 
$R_{dm}= 12 [1+0.07(R_0-8~{\rm kpc})]~{\rm kpc}$~\cite{Freud}. 
 
\subsection{The stellar bulge/bar}

The bulge/bar region is not axisymmetric and can be described by: 
\beq
\rho_{bb}(x,y,z)= \rho_{bb}(0) \left[ s_a^{-1.85} \,\exp(-s_a) + \exp\left(-\frac{s_b^2}{2}\right) \right] \,
\label{bb}
\eeq
where 
\beq
s_a^2 =  \frac{q_b^2 (x^2+y^2)+z^2}{z_b^2}              
\qquad \quad   
s_b^4 = \left[ \left(\frac{x}{x_b} \right)^2 + 
\left(\frac{y}{y_b} \right)^2\right]^2 +  \left(\frac{z}{z_b} \right)^4 \,.
\eeq
$x_b,y_b,z_b$, $q_b$ and $ \rho_{bb}(0)$ 
are considered as free parameters of the model. This is the form proposed by Zhao~\cite{Zhao:1995qh}, based on COBE photometric maps, and it is in fair agreement with the one given by Freudenreich~\cite{Freud}. The second term in Eq.~(\ref{bb}) describes a triaxial bar while the first term represents an axisymmetric nucleus whit a power law behavior 
$\sim s_a^{-1.85}$~\cite{Klypin}. 

In the analysis, we will implement an axisymmetrized version of Eq.~(\ref{bb}), and assume
$x_b \simeq y_b = 0.9~{\rm kpc} \cdot (8~{\rm kpc}/R_0)$, $z_b=0.4~{\rm kpc} \cdot (8~{\rm kpc}/R_0)$ 
and $q_b= 0.6$. These (axisymmetrized) values are in agreement for the best fit obtained in 
Ref.~\cite{Zhao:1995qh} assuming $R_0=8~{\rm kpc}$, and then scaled to an arbitrary $R_0$. 
The choice of fixing these parameters is again related to the lack of observables, among 
those implemented, to discriminate among these values 
and small deviations around them. The normalization $\rho_{bb}(0)$, or equivalently the mass of
the bulge/bar system at a fixed $R_0$, is kept as real free parameter.

\subsection{The dust layer}
The distribution of the interstellar medium is assumed to be axisimmetric
as well. We infer the average density per annulus around the galactic
center of atomic and molecular hydrogen
from the analysis of Dame \cite{dame}, correcting then for helium. The vertical 
distribution is assumed to be the same as for the stellar disc.
The only free parameter entering in the definition of this term is $R_0$, 
as we rescale again with ${R_0}/8~{\rm kpc}$ all radial distances.

\subsection{The dark matter halo}

There are a few possible choices here. One possibility is to follow the scheme suggested 
from results of N-body simulations of hierarchical clustering, and sketch the Milky Way as the 
spherically symmetric radial density profile:
\begin{equation}
  \rho_h(r)=\rho^{\prime} f\left(r/a_h\right)\,,
\label{nbody}
\end{equation}
where, according to the numerical simulations,  $f(x)$ is the function which
sets the universal, or nearly-universal, shape of dark matter halos, while $\rho^{\prime}$ 
and $a_h$ are a mass normalization and a length scale, usually given in terms of the virial mass 
$M_{vir}$ and a concentration parameter $c_{vir}$. We will adopt here the definitions: 
$M_{vir}\equiv 4\pi/3 \Delta_{vir} \bar{\rho}_0\,
R_{vir}^3$, with $\Delta_{vir}$ the virial overdensity
as computed in Ref.~\cite{BN}, $\bar{\rho}_0$ the mean background density
and $R_{vir}$ the virial radius; and $c_{vir} \equiv R_{vir}/r_{-2}$, with
$r_{-2}$ the radius at which the effective logarithmic slope of the profile is $-2$.
The latest simulations favor the Einasto 
profile~\cite{n04,graham}:
\begin{equation}
  f_{E}(x) = \exp\left[-\frac{2}{\alpha_E} \left(x^{\alpha_E}-1\right)\right]\,, 
\label{eq:einasto}
\end{equation}
with the Einasto index $\alpha_E$ ranging about 0.1 to 0.25  (reference value, say,  
$\alpha_E=0.17$). We will also check results for the profile originally proposed by Navarro, 
Frenk and White~\cite{NFW} i.e. 
\beq
f_{NFW}(x)=\frac{1}{x(1+x)^2} \,, 
\eeq
which is most often used. 

While the NFW profile has a $1/r$ singularity towards the center of the Galaxy, this is smoothed 
or partially smoothed by the Einasto index $\alpha_E$. Once more, our analysis is not focussed 
on the central region of the Galaxy, with the model and the implemented constraints that are 
not accurate enough at small radii. We also have a reduced discrimination power with
respect to models for which the central dark matter enhancement is totally erased, such as the
Burkert profile~\cite{Burkert:1995yz}: 
\beq
f_{B}(x)=\frac{1}{(1+x)(1+x^2)} \,.  
\eeq
This model has a density profile with a constant core rather 
than a dark matter cusp, and is phenomenologically motivated since it fits better than the NFW
profile the gentle rise in the rotation curve at small radii which seems to be observed for many 
external galaxies, especially in case of low-mass dark-matter-dominated low surface brightness 
and dwarf galaxies\cite{Gentile}. Another issue which we choose not to address in the present version of the
analysis is the fact that, in principle, one should refer to Eq.~(\ref{nbody}) as to the dark matter 
density profile prior the baryon infall, and then model and take into account the feedback from 
the baryon settling into the center of the Galaxy. Actually how this happens is still a matter of 
debate. The simplest scenario is the one in which the baryon infall could have occurred 
adiabatically, as a smooth and slow process with no net transfer of angular momentum between components; in the approximation of a spherical system in which the local velocity distribution is unchanged as well (as it happens, e.g., if all particles are placed initially on circular orbits),
the adiabatic invariants drive a rather sharp enhancement of the dark matter
concentration in the central region of the Galaxy~\cite{blumental}. At the other extreme, the 
baryon infall could have happened with a sensible exchanged of angular momentum between 
baryons and dark matter, with an equilibrium configurations of the dark matter that
would resemble the Burkert profile~\cite{elzant}. We choose to avoid these issues and assume 
that the net effect from baryons is at most to change the initial concentration parameter to some 
different final concentration parameter, which is the quantity relevant for dynamical observations 
and dark matter detection (this in turns weakens the link between results presented here and   
the correlation pattern between $M_{vir}$ and  $c_{vir}$ seen in N-body simulations).

\section{The Likelihood function}
\label{Likelihood}

The Galactic model of section \ref{GM} is fully specified only when the values of its parameters are fixed. Determining these parameters from the data by statistical inference is essential to investigate halo properties such as the local dark matter density. This task requires a link between the data and the model parameters which, in the present analysis, is established by the Likelihood function.
We summarize briefly below the key concepts and the steps that are needed to derive our results. (For a more detailed description, see, {\it e.g.} \cite{book})

The Likelihood is the joint probability density function (pdf) to observe the whole set of actual data given a fully specified model. It is usually denoted by $\mathcal{L}(d|g)$, where $d$ and $g$ represent the data and the parameters respectively. In the statistical inference the parameters $g$ can be treated either as deterministic variables or as stochastic variables. In the first case the aim of the data analysis is to estimate the model parameters through a maximum Likelihood approach. Although commonly used in the study of the Milky Way, this technique involves a numerical maximization of Likelihood functions and becomes too involved when these are defined in a parameter space of large dimensionality. In the second case, instead, the analysis aims to infer probability regions in the parameter space of the underlying Galactic Model. It turns out that this second approach is the most appropriate when the Likelihood is a complicated function of a large number of parameters.   

Seen as a function of the model parameters, the Likelihood is not a true pdf defined in the parameter space. It is however mapped into a pdf for the model parameters by means of Bayes' theorem, which states      
\beq
p(g|d) = \frac{\mathcal{L}(d|g) p(g)}{p(d)} \,,
\label{Bayes}
\eeq
where p(d) is the Bayesian evidence and in the present discussion simply plays the r\'ole of a normalization constant. The function $p(g)$, instead, is the prior probability density function and encodes our prejudice concerning the most probable values for the parameters before seeing the data. The prior distribution can be crucial or irrelevant in the inference procedure depending on how peaked is the Likelihood and therefore on how many informations are carried by the used data. 
The pdf $p(g|d)$ is called the posterior probability density function and it is the main target of every Bayesian probabilistic inference. It reflects the change of our prejudice about the most probable values for the model parameters after seeing the data. When the posterior pdf has been determined, then one can: 
\begin{itemize}
\item construct marginal posterior pdf integrating $p(g|d)$ over portions of the parameter space;
\item infer posterior pdf for generic functions $f$ of the model parameters;
\item calculate averages and variances of the desired quantities with respect to $p(g|d)$.
\end{itemize}

Given a n-dimensional parameter space with parameters denoted by $g=g_1,\dots,g_n$, the marginal posterior pdf for $g_1$ reads
\beq
p(g_1|d) = \int dg_{2}\dots dg_{n}\,p(g|d) \,.   
\eeq 
Analogous expressions hold for portions of the parameter space of dimension larger then one. In the next section, for instance, we will calculate two dimensional marginal posterior pdf to look for correlations between different quantities of interest.
The posterior pdf for any function $f$ of the model parameters is denoted by $p(f,g|d)$. It is related to the posterior pdf $p(g|d)$ by the relation
\beq
p(f,g|d) = \delta(f(g)-f)\,p(g|d) \,,
\label{fg}
\eeq
which follows from the definition of conditional pdf. In practise, as explained in the appendix, $f$ and $g$ are discrete variables varying within a given sample. In this case, the delta-function in Eq.~(\ref{fg})is equal to one for $f=f(g)$ and zero otherwise. 
Finally, expectation values of any function $f$ of the model parameters with respect to $p(g|d)$ can be calculated as follows  
\beq
\langle f(g) \rangle = \int dg \,f(g)\,p(g|d) \,.   
\label{ev}
\eeq
A subset of the functions $f$ considered in the present paper is given in tables \ref{table_einasto} and \ref{table_nfw}. We will call them, for obvious reasons, derived quantities. A particular emphasis will be attached to the case $f=\rho_{\rm DM}(R_0)$ for its relevance in the physics of direct dark matter detection. Let us also mention, that in the numerical calculations, we evaluated the integrals introduced in this section using Markov Chain Monte Carlo methods. Further details on this approach can be found in the Appendix. 

We conclude this section giving an explicit expression for the Likelihood function used in our analysis. Different datasets will be considered as statistically independent, which means that the total Likelihood is given by the product of the single Likelihoods obtained from each set independently.  
We start by introducing the 7-dimensional array: 
\beq
D^j = (A-B,\,A+B,\,v_c(R_0),\,\Sigma_{|z|<1.1\,\rm kpc},\, \Sigma_{*},\,M(<50 \rm kpc), \,M(<100 \rm kpc)) \,\, 
\eeq
which has in the entries the experimental data of section \ref{data}. To each entry of $D^{j}$ one can associate a theoretical prediction $T^j$ and an error $\sigma^j$ as explained in the previous sections. The terminal velocities $v_{\rm ter}^{i}$,  the velocity dispersions in a tracer of population $\sigma_r^{*\,i}$, and finally, the proper motions $(\mu_{l}^{i},\mu_{b}^{i})$, distances $d^{i}$ and local standard of rest velocities of high mass star forming regions $v_{\rm lsr}^{i}$, will be treated separately (here $i=1,\dots,m$ with $m=111$ for the terminal velocities data, 9 for the velocity dispersions data and 11 for the high mass star forming regions data). Errors associated to these datasets are denoted by $\sigma_{\rm ter}^{i}$, $\sigma_{\rm r}^{i}$, $\sigma_{\rm d}^{i}$,  $\sigma_{\rm l}^{i}$, $\sigma_{\rm b}^{i}$and $\sigma_{\rm lsr}^{i}$, respectively. If a given quantity carries a bar ({\it e.g.} $\bar{v}_{\rm ter}^{i}$), it denotes the experimental value of the corresponding quantity. The same symbol without a bar ({\it e.g.} $v_{\rm ter}^{i}$) represents the theoretical prediction for such a quantity. With this notation we can now write the total Likelihood as follows
\beqra
-\ln \mathcal{L} &=& \frac{1}{2}\sum_{j=1}^{7}\frac{(T^j-D^j)^2}{\sigma^{j\,2}} + 
\frac{1}{2}\frac{1}{111}\sum_{i=1}^{111}\frac{(v_{\rm ter}^{i}-\bar{v}_{\rm ter}^{i})^2}{\sigma_{\rm ter}^{i\,2}} +
\frac{1}{2}\frac{1}{9}\sum_{i=1}^{9}\frac{(\sigma_{r}^{*\,i}-\bar{\sigma}^{*\,i}_{r})^2}{\sigma_{r}^{i\,2}}\nonumber\\
&+&\frac{1}{2}\frac{1}{44}\sum_{i=1}^{11}\min_{R>R_0} 
\Bigg[ \frac{(d^{i}(R)-\bar{d}^{i})^2}{\sigma_{\rm d}^{i\,2}}+ 
\frac{(\mu_{\rm l}^{i}(R)-\bar{\mu}_{\rm l}^{i})^2}{\sigma_{\rm l}^{i\,2}} + 
\frac{(\mu_{\rm b}^{i}(R)-\bar{\mu}_{\rm b}^{i})^2}{\sigma_{\rm b}^{i\,2}} \nonumber\\      
&+& \frac{(v_{\rm lsr}^{i}(R)-\bar{v}_{\rm lsr}^{i})^2}{\sigma_{\rm lsr}^{i\,2}}    \Bigg] \,. 
\label{Like}
\eeqra
For data referring to the high mass star forming regions, we have introduced a minimization procedure 
since velocities and distances have high correlated errors and values of $R$ are not univocally 
defined (see the discussion in \cite{Dehnen:1996fa}).

Concerning the choice of the prior pdf, we assign flat priors to all the model parameters. We can therefore write $p(g) = \prod_{k=1}^{n} p_{k}(g_k)$ where the single $p_{k}(g_k)$ read
\beq
p_{k}(g_k) = \left\{ \begin{array}{lr} 
{\rm const} & {\rm for} \,\, g_k^{\rm min} < g_k < g_k^{\rm max} \\
&\\
0 & {\rm otherwise}
\end{array}
\right.
\eeq
The ranges $(g^{\rm min}_k,g^{\rm max}_k)$ for the parameters of the underlying Galactic model are listed in table \ref{priors}.

\section{Results}
\label{res}

We can now present the results of our analysis, namely the marginal posterior pdf, means, standard deviations and confidence intervals for different quantities of interest. As already stressed in the previous section, a particular emphasis will be given to the local dark matter density, since it is a very important quantity for direct dark matter detection experiments.

To reach this goal, we run a dedicated FORTRAN code which is partially based on Superbayes \cite{superbayes} and cosmomc \cite{cosmomc}, from which we borrowed Markov Chain Monte Carlo routines.

In the analysis we consider separately different functional forms for the dark matter halo density; we start with the case of the Einasto and NFW profiles. The latter has been the most commonly used so far in the literature, while the former seems to be favored from the most recent N-body simulations. As we will show, however, results are essentially independent from the choice of the dark matter profile, since as we already stressed the two profiles mainly differ in the central portion of the Galaxy where our set of dynamical constraints are less efficient. 

\begin{figure}
\begin{center}
\includegraphics[width=7.8 cm]{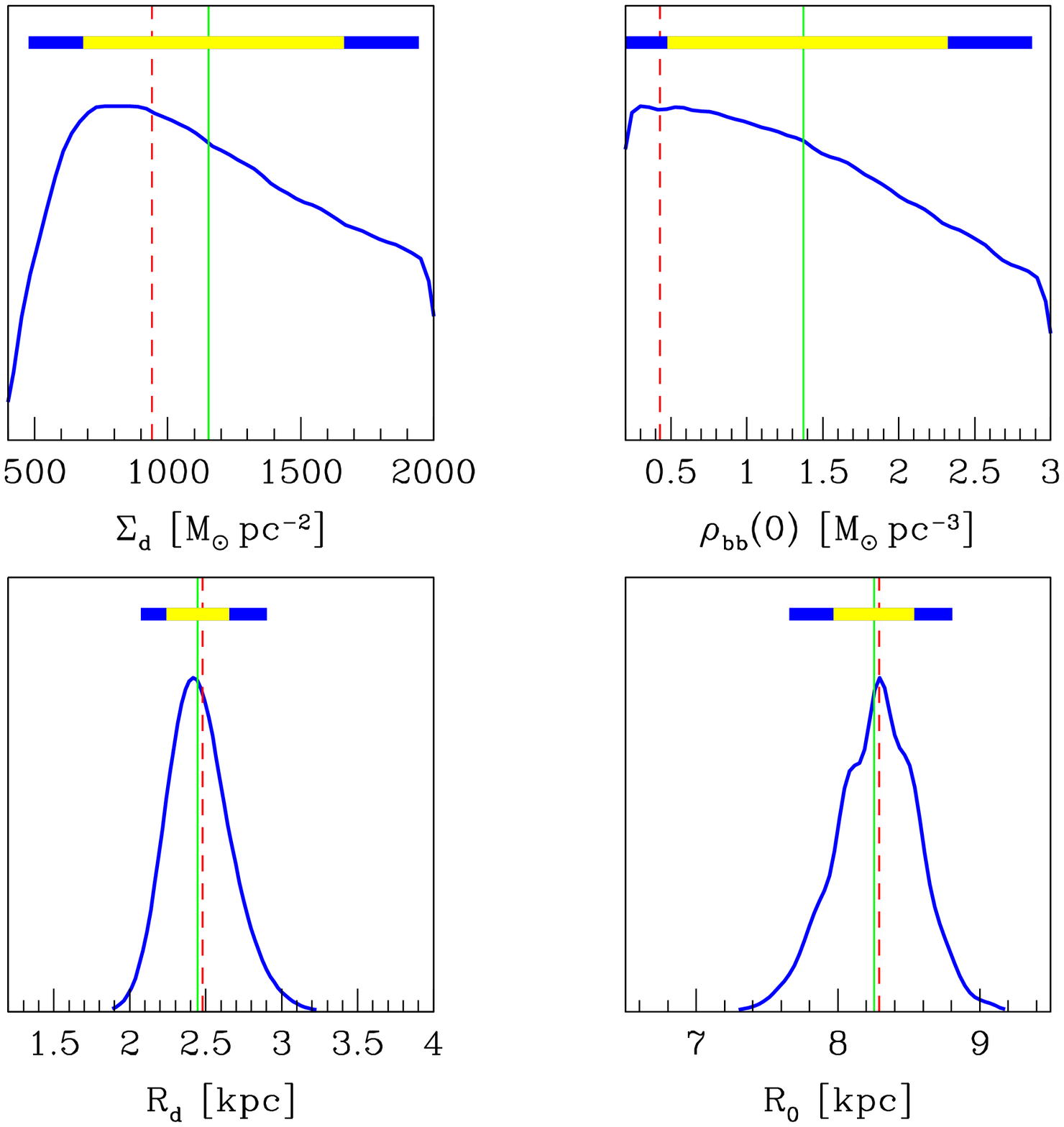}
\includegraphics[width=7.8 cm]{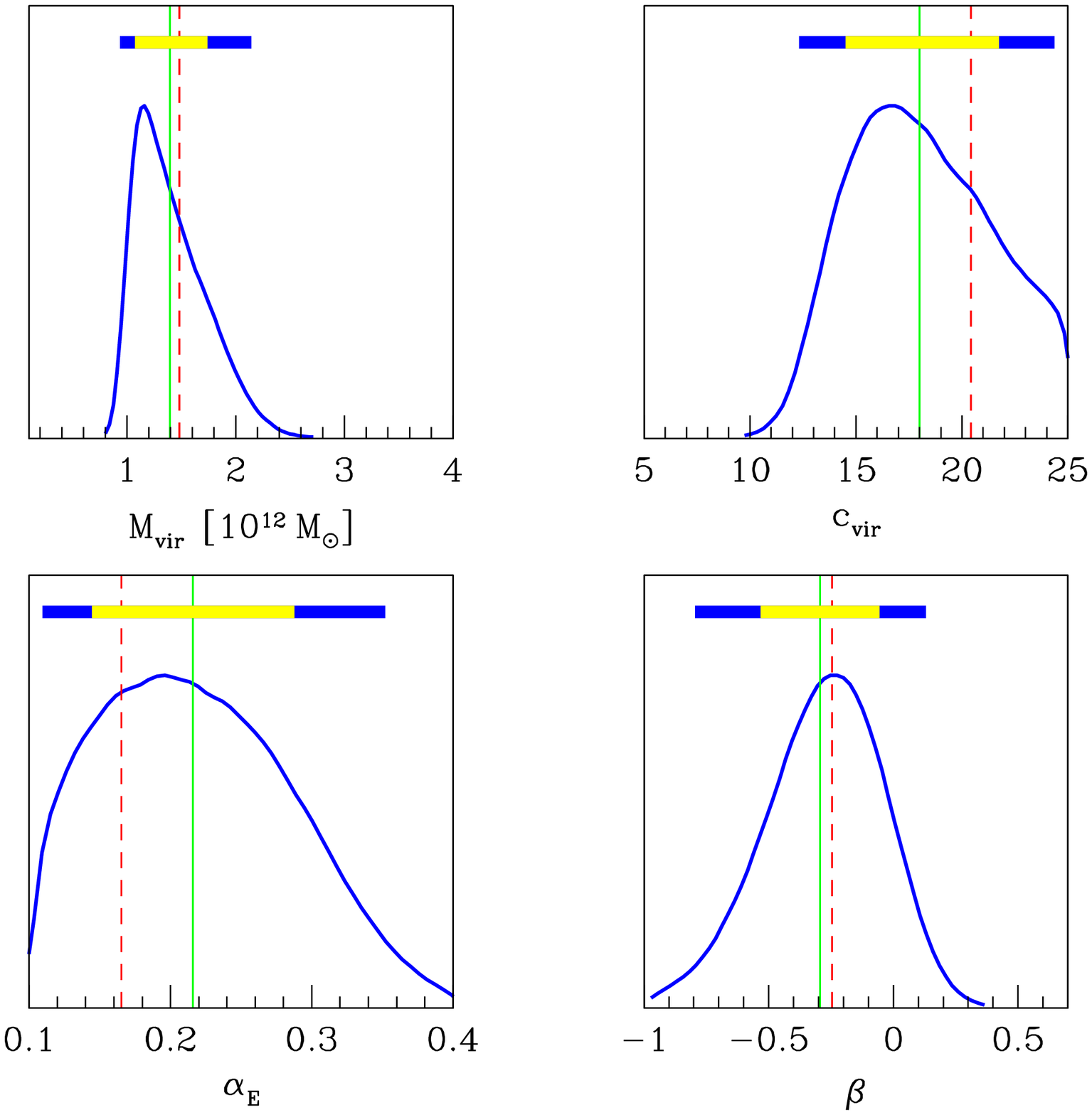}
\caption{\label{param1D_einasto}  Marginal posterior pdf of the Galactic model parameters (Einasto profile). In each plot the green vertical line represents the mean of the corresponding parameter while the red dashed line its best fit value. The yellow (blue) bar above the curves indicates the largest interval including the 68\% (95\%) of the total probability.}
\end{center}
\end{figure}
\begin{figure}
\begin{center}
\includegraphics[width=7.8 cm]{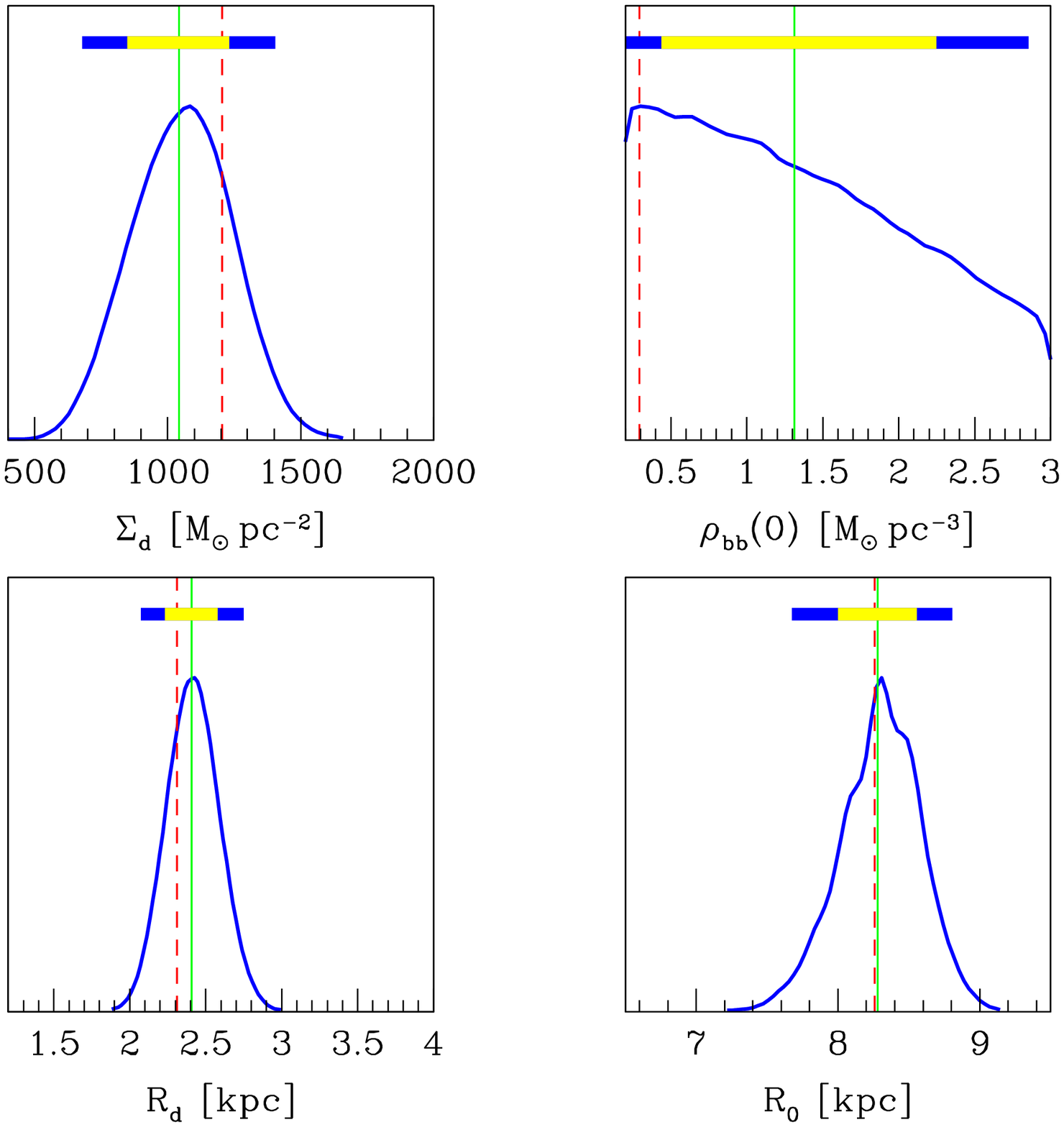}
\includegraphics[width=7.8 cm]{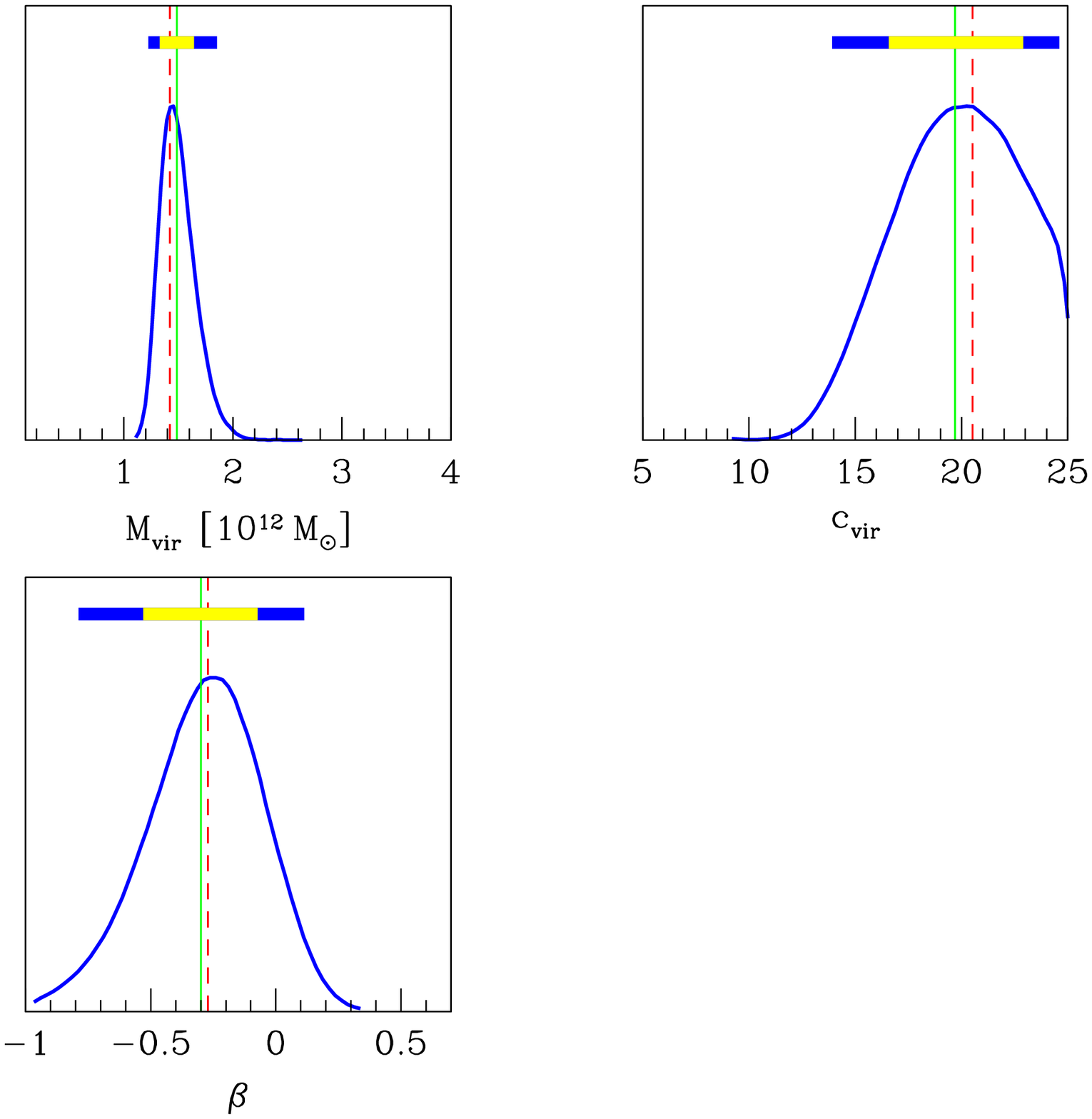}
\caption{\label{param1D_nfw}   Marginal posterior pdf of the Galactic model parameters (NFW profile). In each plot the green vertical line represents the mean of the corresponding parameter while the red dashed line its best fit value. The yellow (blue) bar above the curves indicates the largest interval including the 68\% (95\%) of the total probability.}
\end{center}
\end{figure}
Fig.~(\ref{param1D_einasto}) shows the pdf of the Galactic model parameters when an Einasto profile is assumed. Except for the central bulge/bar energy density $\rho_{bb}(0)$ and the central surface mass density of the disc $\Sigma_d$, which are poorly constrained, the other parameters exhibit a clear peaked distribution. In particular the virialization mass $M_{vir}$ is very well constrained within our dataset. Analogously, Fig.~(\ref{param1D_nfw}) shows the pdf for the model parameters in case of a NFW profile. There is a clear peaked distribution for all the dark matter related parameters as well as now for both disc parameters. 
\begin{figure}
\begin{center}
\includegraphics[width=7.8 cm]{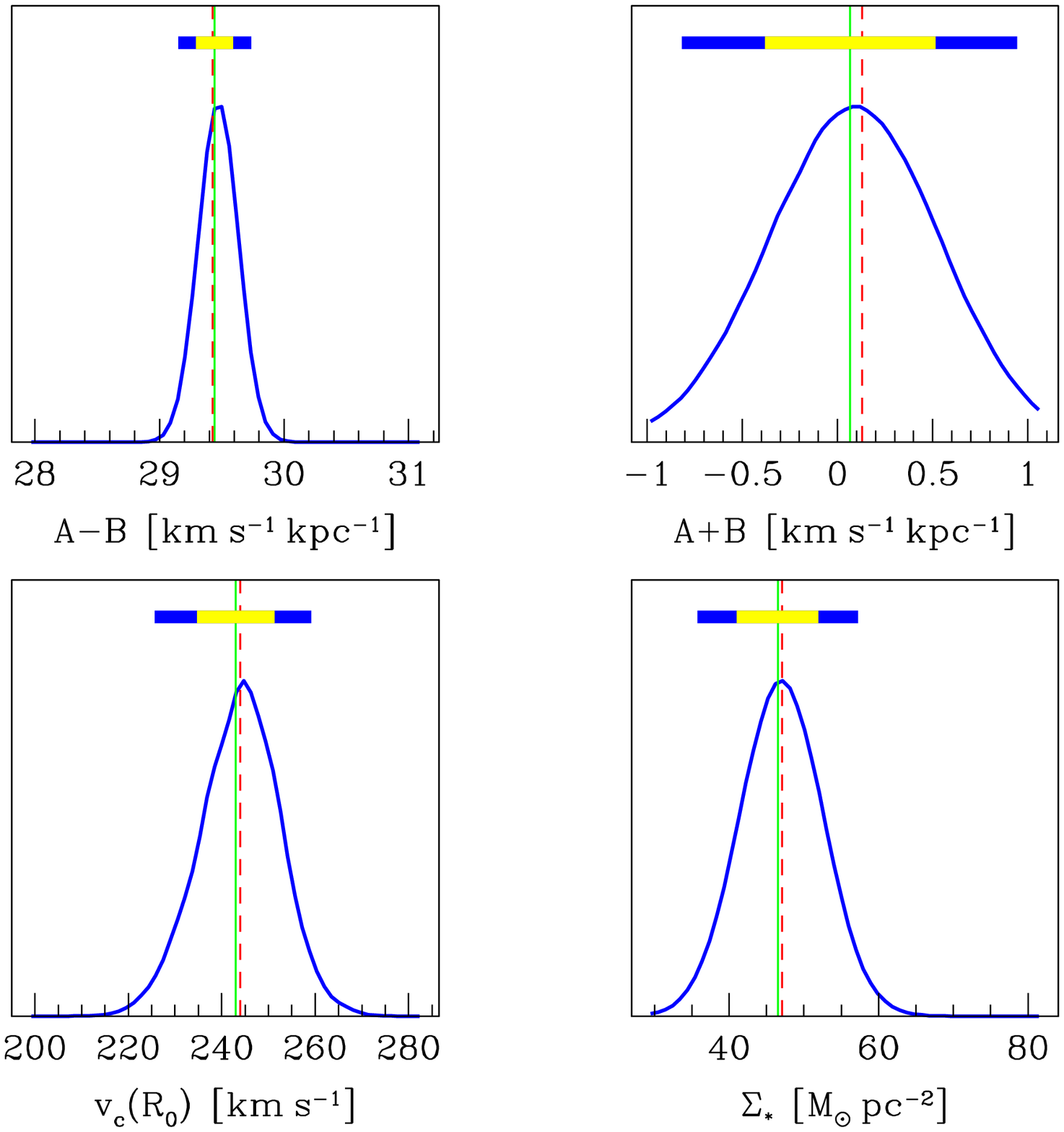}
\includegraphics[width=7.8 cm]{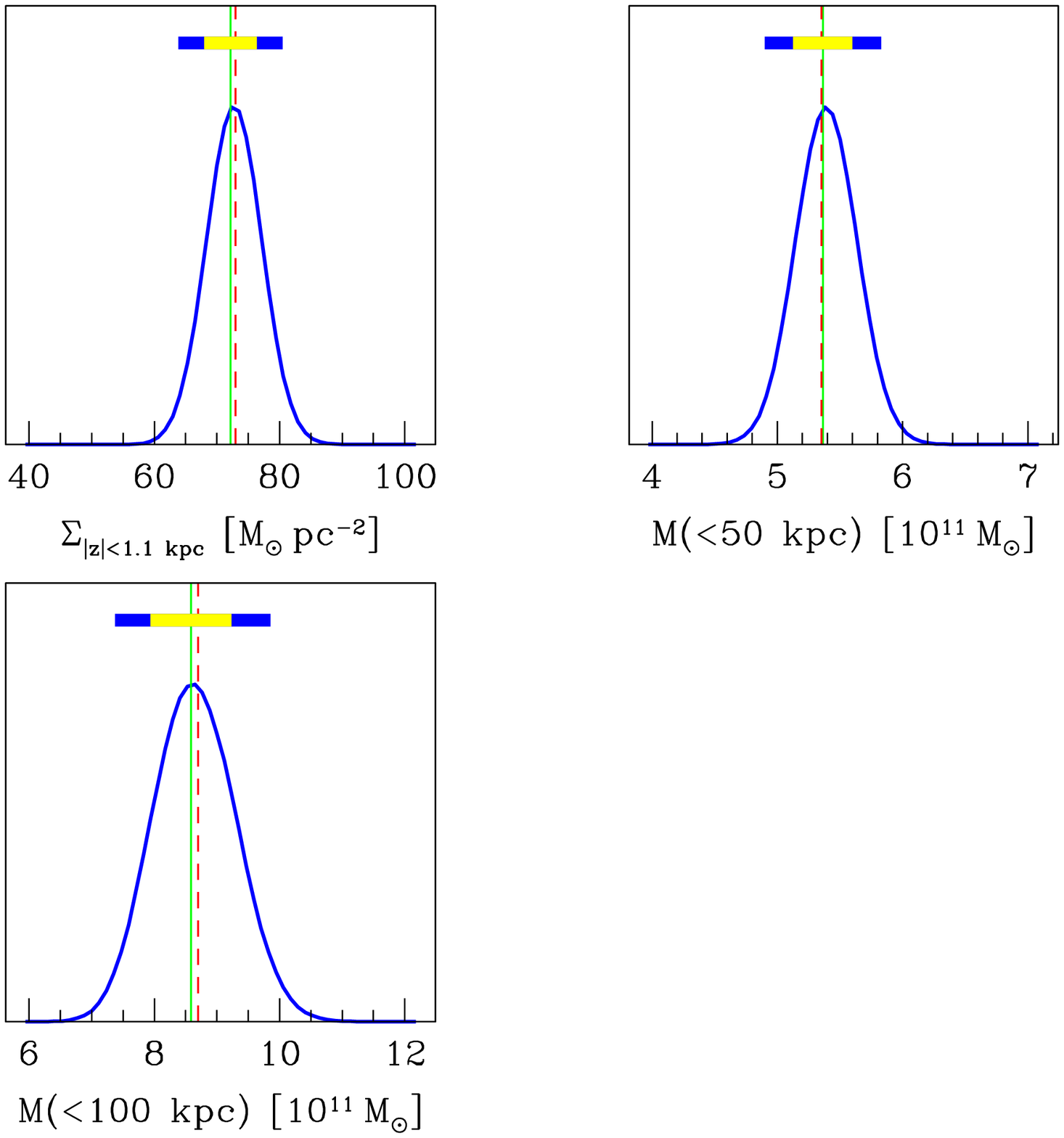}
\caption{\label{deriv1D_einasto}  Marginal posterior pdf of a few derived quantities (Einasto profile). In each plot the green vertical line represents the mean of the corresponding parameter while the red dashed line its best fit value. The yellow (blue) bar above the curves indicates the largest interval including the 68\% (95\%) of the total probability.}
\end{center}
\end{figure}
Fig.~(\ref{deriv1D_einasto}) shows marginal posterior pdf for a few derived quantities which are some of the quantities we implemented as inputs in the analysis. The figure clearly shows consistency between input and output in our procedure. Note in particular that the local circular velocity, which is a quantity entering critically in direct detection experiments, is rather sharply peaked around its central value of 243~km~s$^{-1}$. 
\begin{figure}
\centering
\includegraphics[width=1.\textwidth]{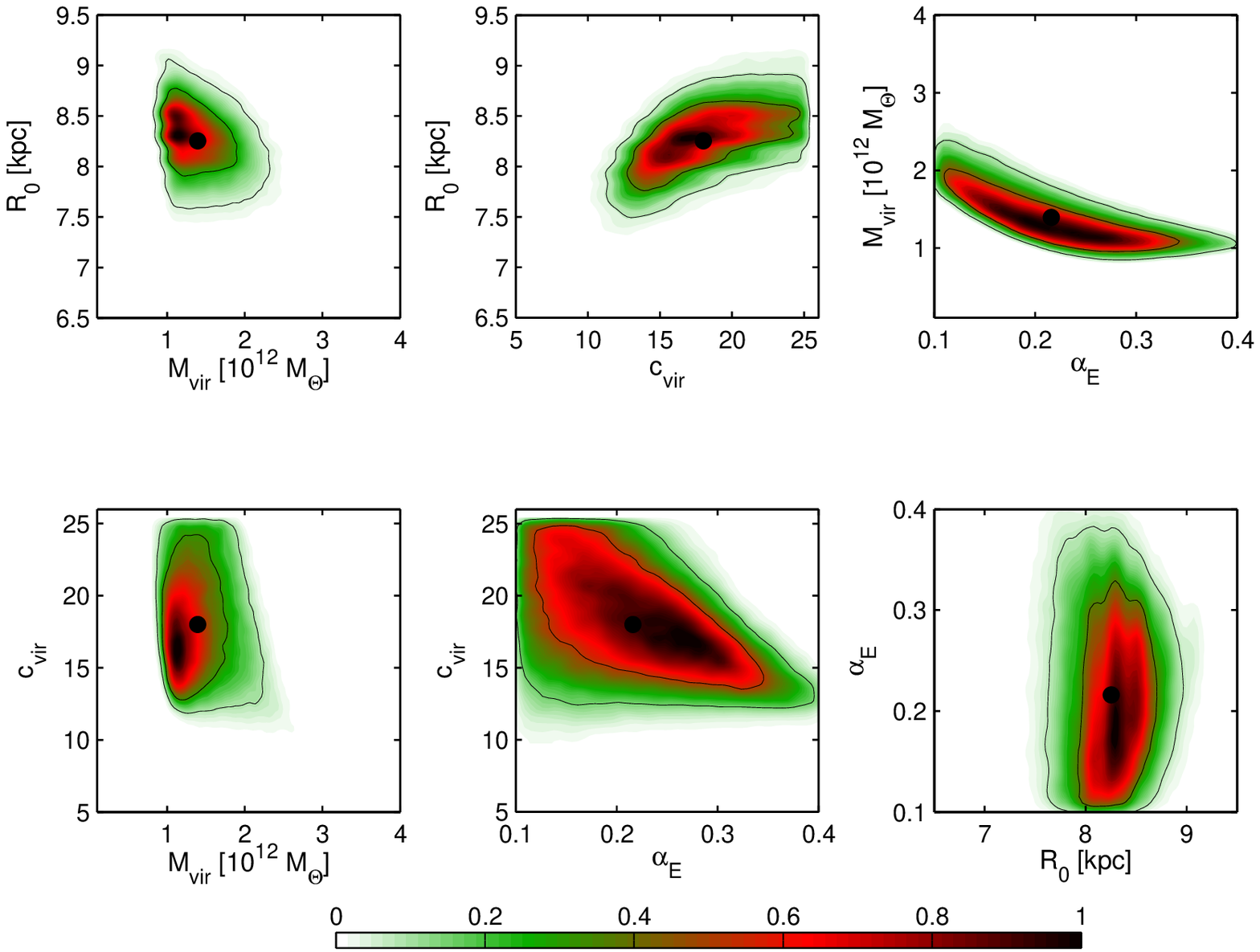}
\caption{\label{param2d1} Two dimensional marginal posterior pdf in the planes spanned by the combinations of the Galactic model parameters which determine the dark matter halo in the case of an Einasto profile. The normalization is such that at the maximum the posterior pdf is equal to one. The black dots correspond to the means of the plotted posterior pdf. One and two sigma contours are also shown.}
\end{figure}
\begin{figure}
\centering
\includegraphics[width=1.\textwidth]{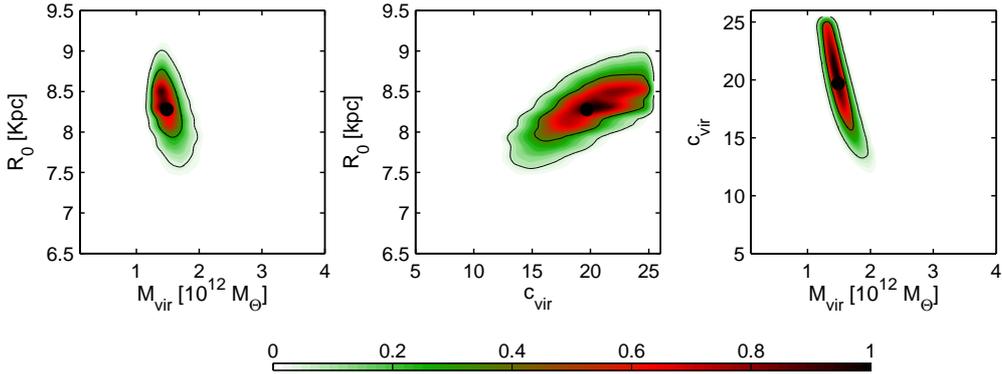}
\caption{\label{param2d2} Two dimensional marginal posterior pdf in the planes spanned by the combinations of the Galactic model parameters which determine the dark matter halo in the case of a NFW profile. The normalization is such that at the maximum the posterior pdf is equal to one. The black dots correspond to the means of the plotted posterior pdf. One and two sigma contours are also shown.}
\end{figure}

From Figs.~(\ref{param2d1}) and (\ref{param2d2}), one can see the correlations between the parameters defining, respectively, the Einasto and NFW halo, as well as the local galctocentric distance. In the first case, the Einasto index $\alpha_E$ is poorly constrained, reflecting again our lack of information at small Galactic radii. On the other hand viral mass and concentration parameter are much more efficiently constrained. The mean value of the concentration parameter is about 18 which stands at the upper end of the value that is predicted by N-body simulation for a dark matter halo of virial mass around $10^{12} M_{\odot}$; note however the possible mismatch between our result and N-body simulation results eventually related to the baryon infall effect. One can also see that the concentration parameter is fairly correlated to the value of the local galactocentric distance which in our distribution has a mean of about $8.2$~kpc. Analogously the result for the NFW case gives a very accurate determination of the virial mass and a sharp correlation between virial mass and concentration parameter. The concentration parameter is again large with a mean equal to 19.7.

\begin{figure}
\begin{center}
\includegraphics[width=6.8 cm]{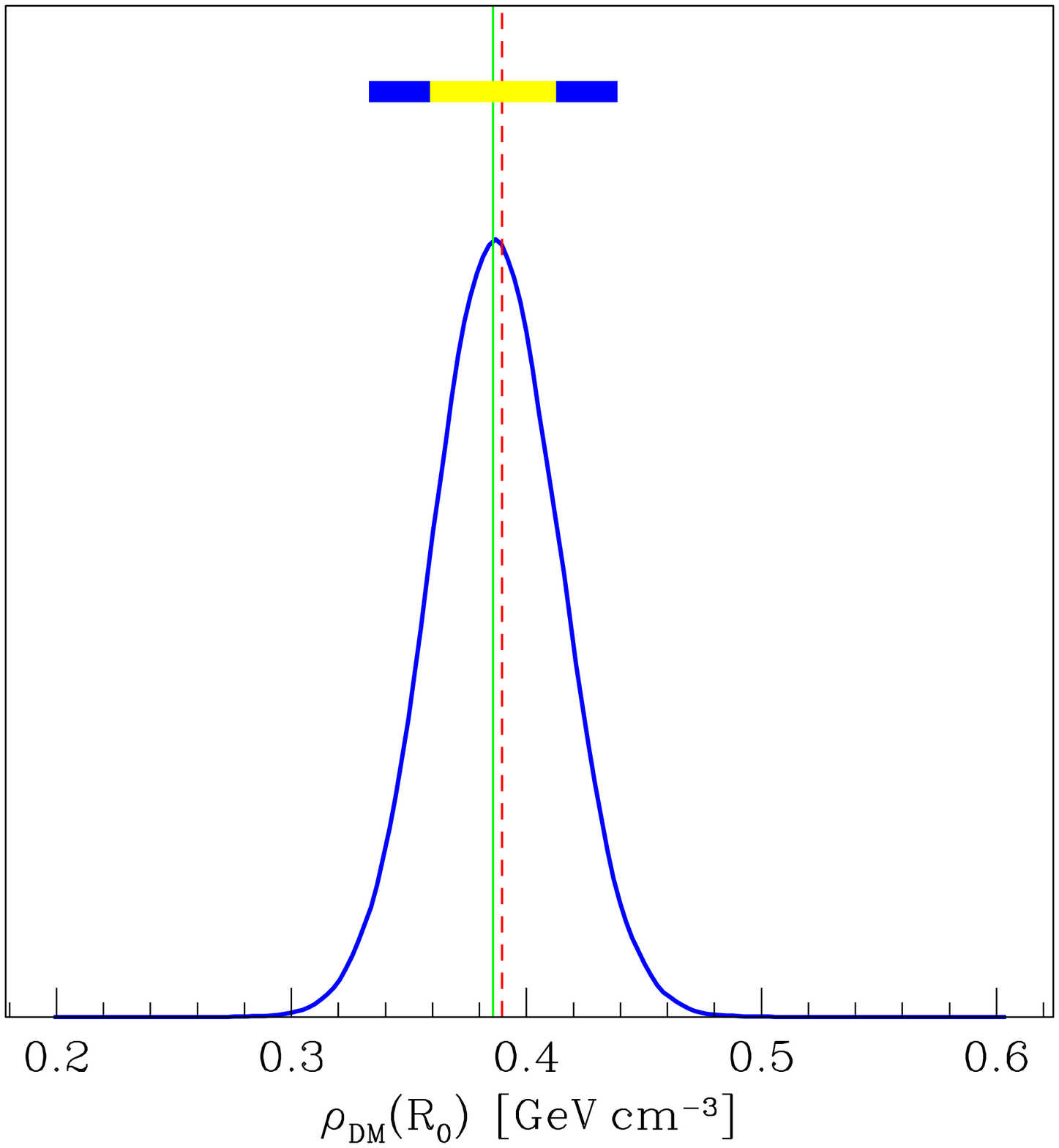}
\includegraphics[width=6.8 cm]{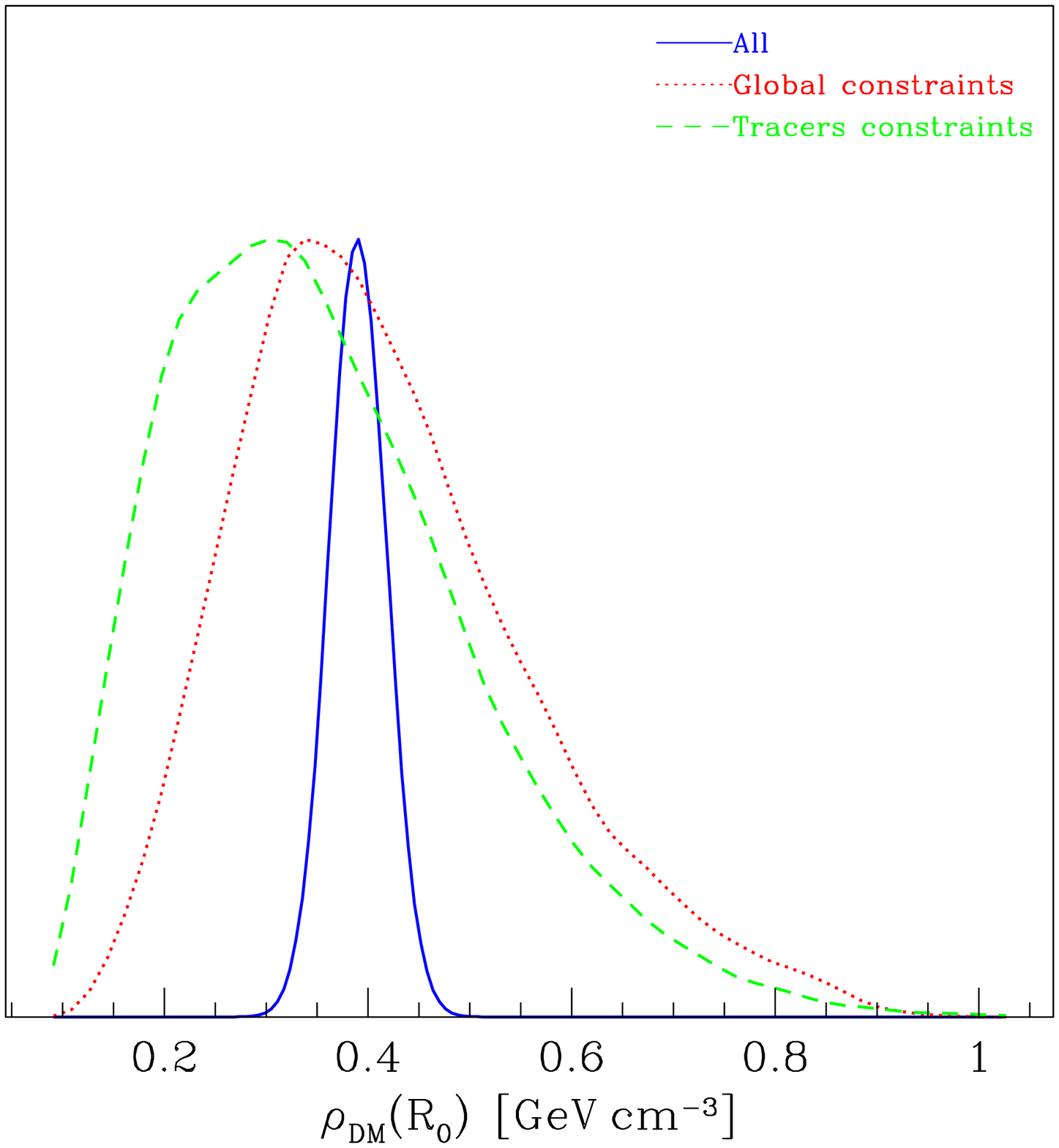}
\includegraphics[width=6.8 cm]{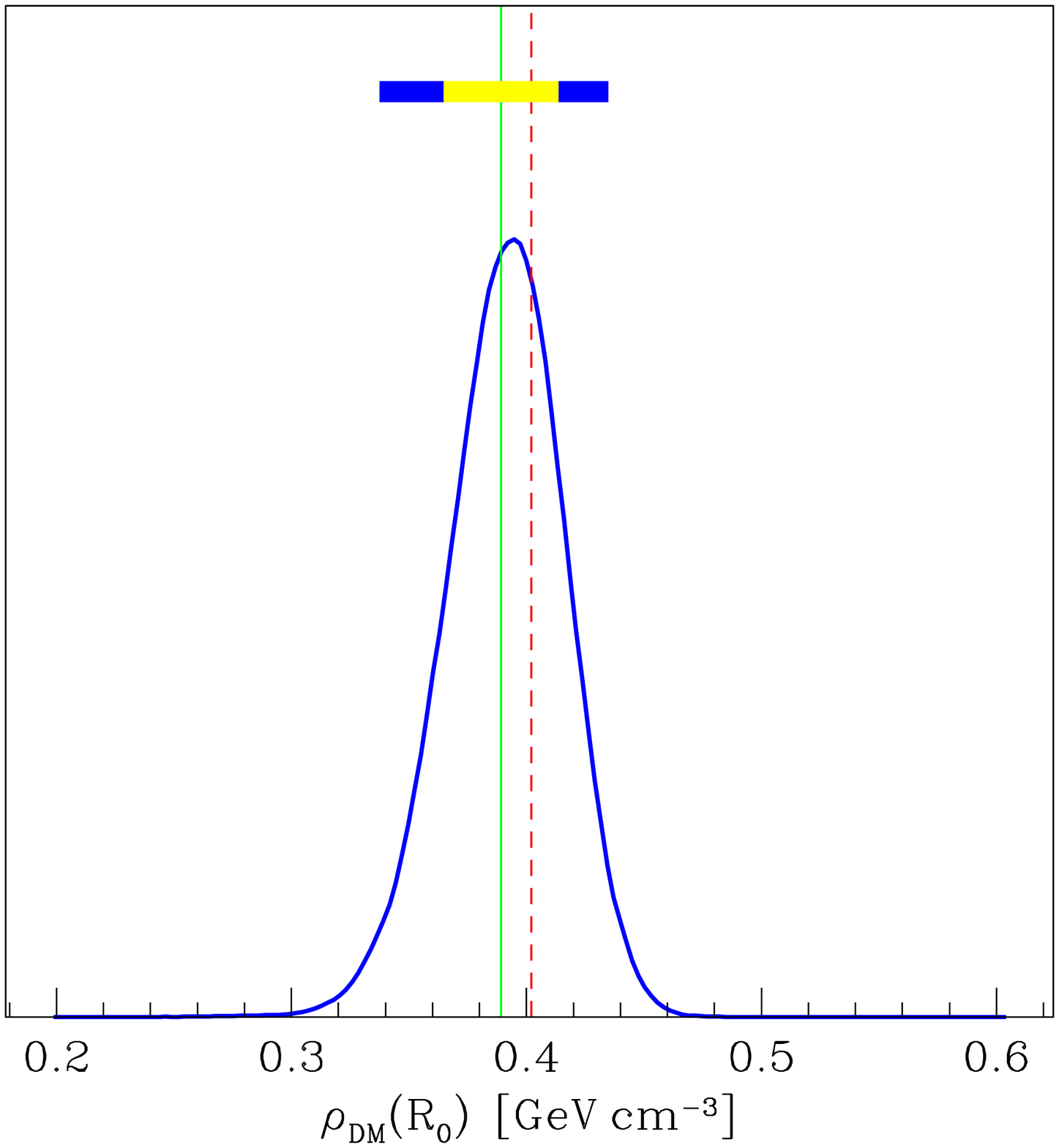}
\includegraphics[width=6.8 cm]{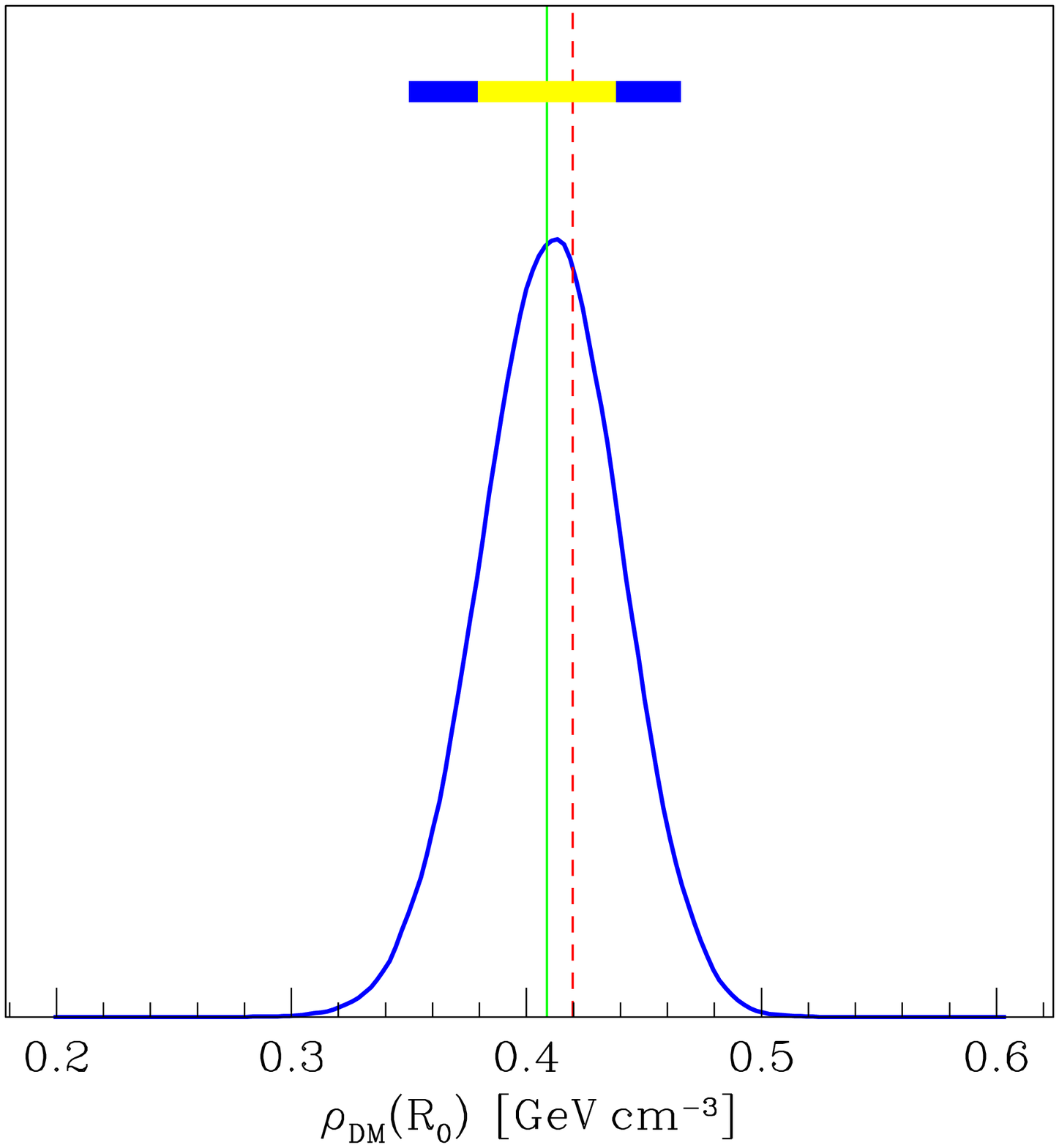}
\caption{\label{rho} Marginal posterior pdf for the local Dark Matter density.Top left panel: assuming an Einasto profile and applying all the constraints. Top right panel: assuming an Einasto profile and applying different subsets of constraints. Global constraints include $M(<50 \rm kpc)$, $M(<100 \rm kpc)$ and $\Sigma_{|z|<1.1 \rm kpc}$. Tracers constraints include the local standard of rest data, the terminal velocities and data referring to the high mass star forming regions. Bottom left panel: assuming a NFW profile and applying all the constraints. Bottom right panel: assuming a Burkert profile and applying all the constraints. Curves and bars have the same meaning as in the previous plots.}
\end{center}
\end{figure}
Fig.~(\ref{rho}) contains the main result in this analysis. We show the marginal posterior pdf for the local dark matter density $\rho_{DM}(R_0)$ for the Einasto profile (top left panel) and the NFW profile (bottom left panel). In the same plot we also show the result obtained when modelling the dark matter halo with the Burkert profile (bottom right panel), in analogy to the analysis just described for the other two profiles. The result is essentially halo profile independent, with the 68\% confidence level associated to this average only slightly changing in the three cases. For the Einasto and NFW profiles $\rho_{DM}(R_0) = 0.385 \pm 0.027\,\rm GeV \,cm^{-3}$ and  $\rho_{DM}(R_0) = 0.389 \pm 0.025\,\rm GeV \,cm^{-3}$, respectively, while in the Burkert case we find $\rho_{DM}(R_0) = 0.409 \pm 0.029\,\rm GeV \,cm^{-3}$. We can therefore give a profile independent determination of $\rho_{DM}(R_0)$ which has an accuracy of roughly the 7\%.

\begin{figure}
\centering
\includegraphics[width=1.\textwidth]{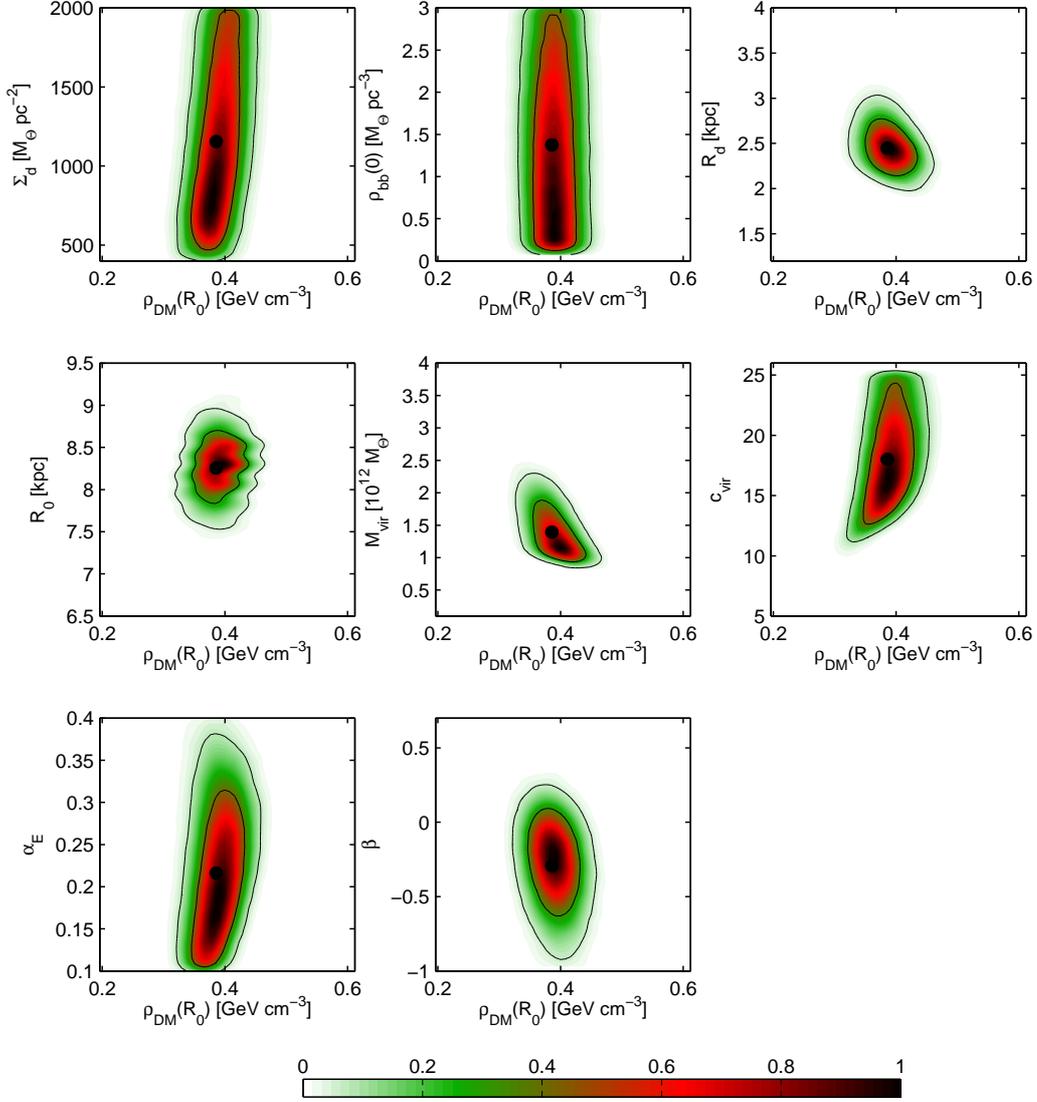}
\caption{\label{rhoparam} Two dimensional marginal posterior pdf in the planes spanned by the local dark matter density and one of the Galactic model parameters in the case of an Einasto profile. The normalization is such that at the maximum the posterior pdf is equal to one. The black dots correspond to the means of the plotted posterior pdf. One and two sigma contours are also shown.}
\end{figure}

\begin{figure}
\centering
\includegraphics[width=1.\textwidth]{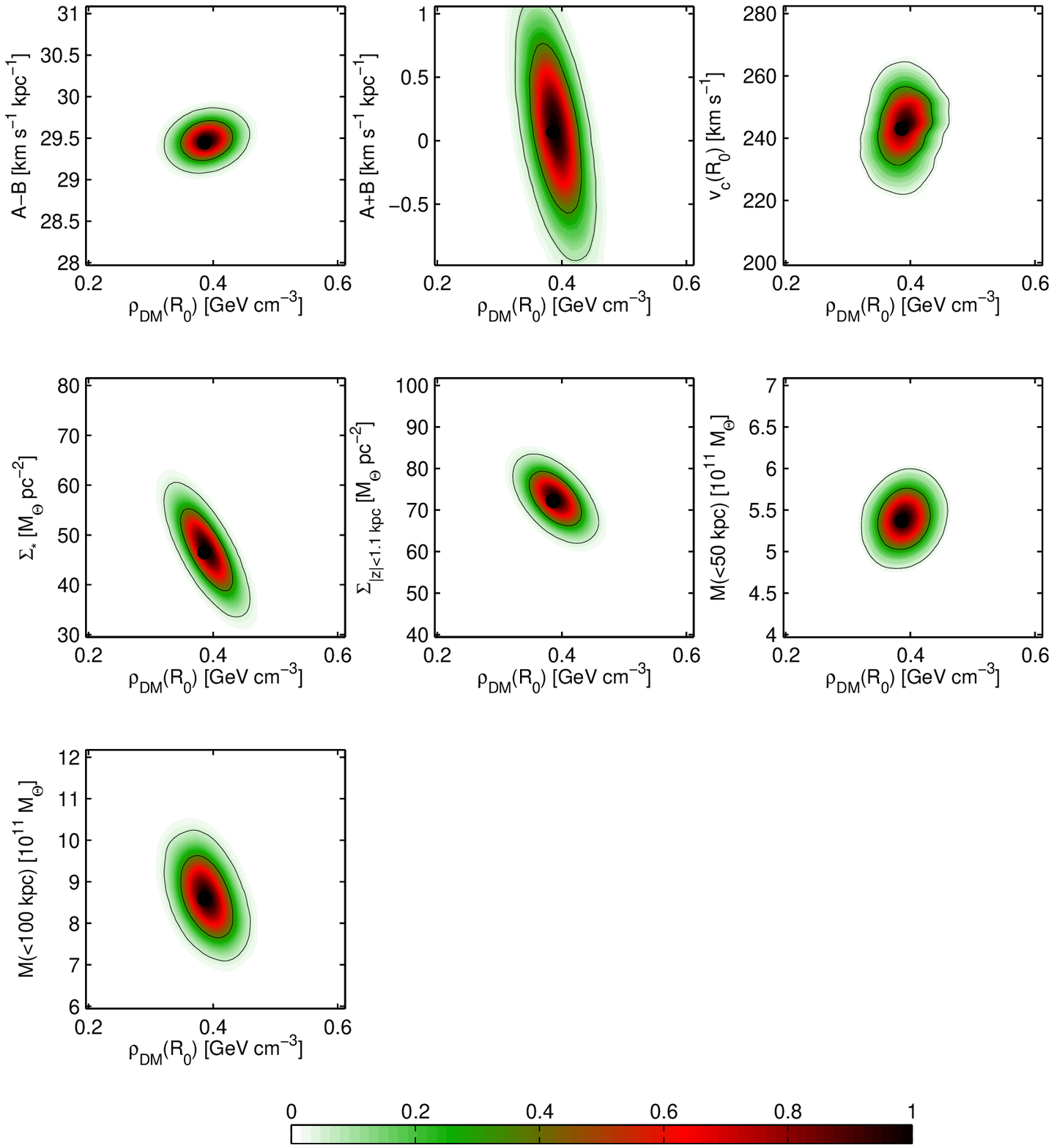}
\caption{\label{rhoderiv} Two dimensional marginal posterior pdf in the planes spanned by the local dark matter density and one of the derived quantities in the case of an Einasto profile. The normalization is such that at the maximum the posterior pdf is equal to one. The black dots correspond to the means of the plotted posterior pdf. One and two sigma contours are also shown.}
\end{figure}
\begin{figure}
\begin{center}
\includegraphics[width=7.5 cm, angle=-90]{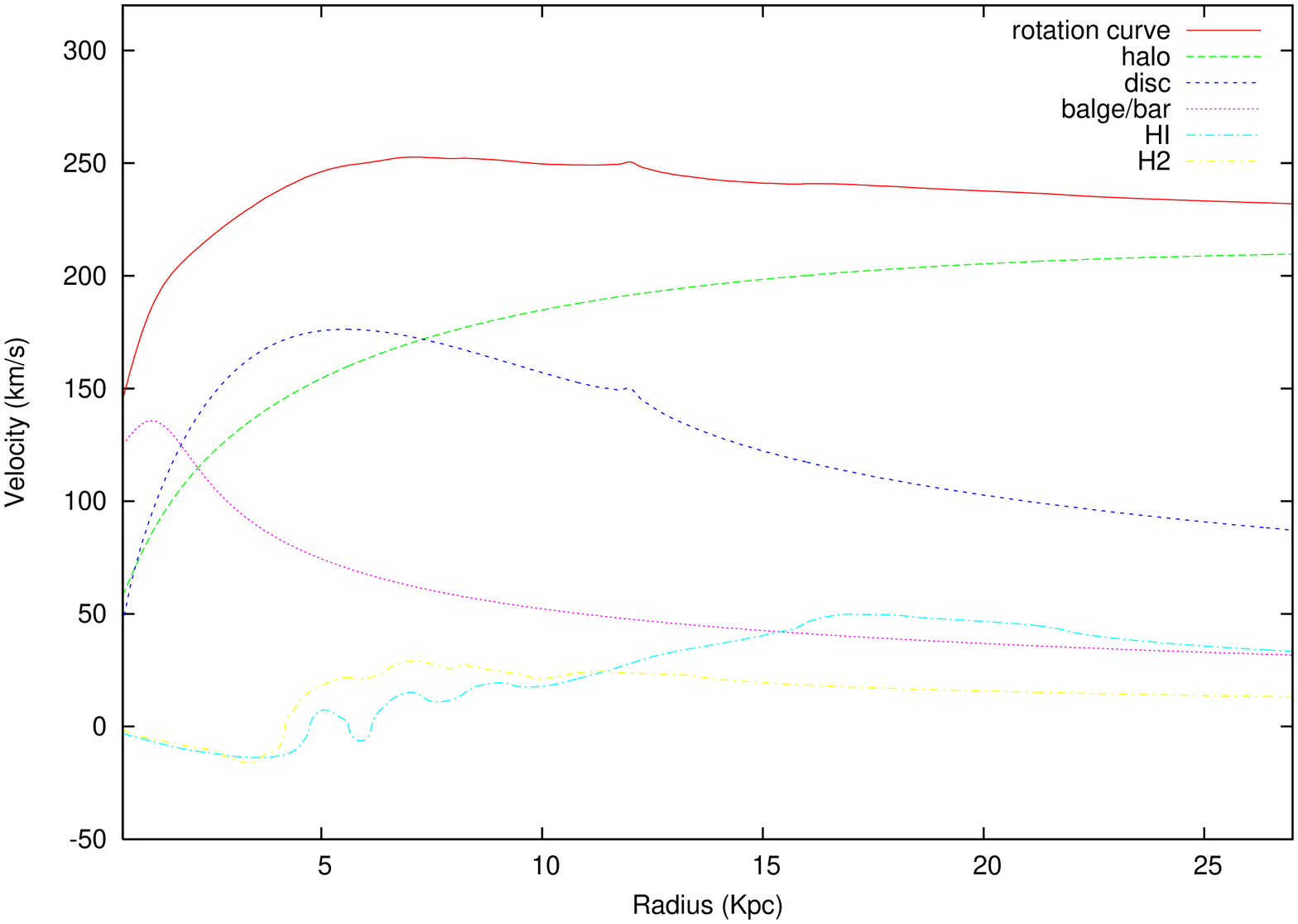}
\caption{\label{rotcurv}Top panel: Galactic rotation curve in the case of an Einasto profile. Different curves are associated to the contributions of the various Galactic components.} 
\end{center}
\end{figure}

Figs.~(\ref{rhoparam}) and~(\ref{rhoderiv}) show two dimensional marginal posterior pdf for  $\rho_{DM}(R_0)$ against all the parameters in the Milky Way mass model and a set of input observational quantities, in case of the Einasto profile. One can deduce that the very precise determination of $\rho_{DM}(R_0)$ is not stemming from any single particular observable or from an ad hoc choice of the parameters defining the mass model since, apart from the obvious correlation with the local mass observables such as the local stellar disc surface density and the local total integrated surface density, other clear correlations are more difficult to single out. On the contrary, the small error on $\rho_{DM}(R_0)$ comes from a combination of complementary constraints. To understand which pieces of information are needed to determine $\rho_{DM}(R_0)$, it is useful to write the local dark matter density as follows
\beq
\rho_{DM}(R_0) = \frac{1}{4\pi G R_0^2} \frac{\partial}{\partial R} \Big( R\Theta^{2} \Big)_{R=R_0} - K \,,
\label{eqrho}
\eeq 
where $K$ includes a contribution to $\rho_{DM}(R_0)$ from the mass density of the baryons plus a correction due to the fact that such a component can not be treated as spherically symmetric. The local amount of baryonic matter is essentially constrained by the local disc surface density $\Sigma_*$. The first term in Eq.~(\ref{eqrho}), instead, depends from several aspects of the Galactic model, namely the Sun's galactocentric distance, the local values of the rotation curve and its local slope. Note that all this informations are mainly associated to local observables. We find that a combination of the precise estimates for $\Theta_0$ and $A-B$ quoted ins section \ref{data}, together with the informations on the slope of the rotation curve coming from $A-B$, the terminal velocities and other tracers, are efficient datasets to constrain $\frac{1}{R_0^2}\frac{\partial}{\partial R}\Big( R\Theta^{2}\Big)_{R=R_0}$. It should be now clear that global constraints alone - such as  $M(<50 \rm kpc)$, $M(<100 \rm kpc)$ and $\Sigma_{|z|<1.1 \rm kpc}$ - or constraints associated to different tracer populations alone, could have not lead to a determination of the local dark matter density as precise as the one we quote in this paper (see top right panel in Fig.~(\ref{rho})).
In Fig.~(\ref{rotcurv}) we plot, for the Einasto profile, the rotation curve calculated for the point in the parameter space corresponding to the mean of the Einasto posterior pdf.

\begin{figure}
\begin{center}
\includegraphics[width=7.8 cm]{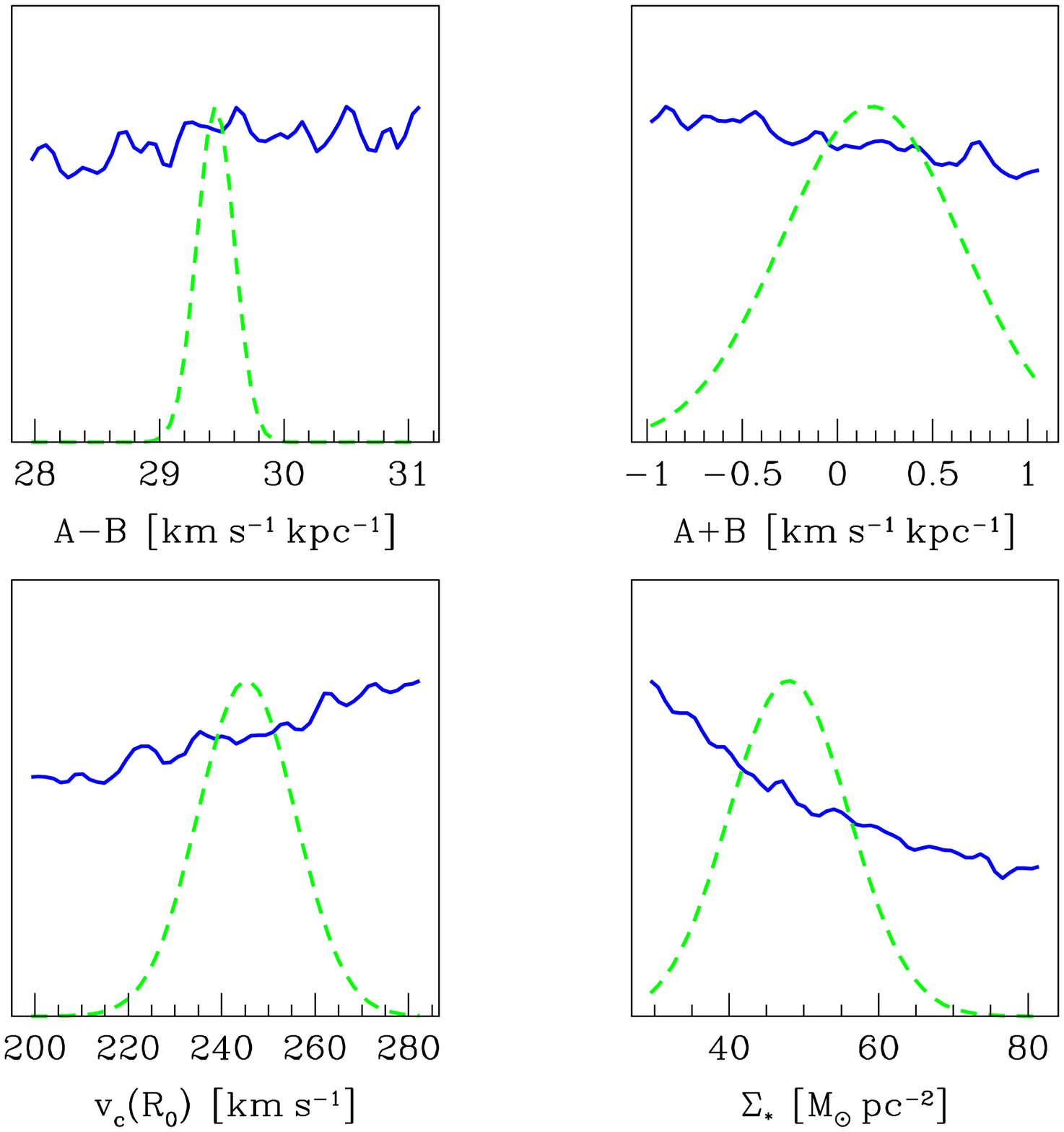}
\includegraphics[width=7.8 cm]{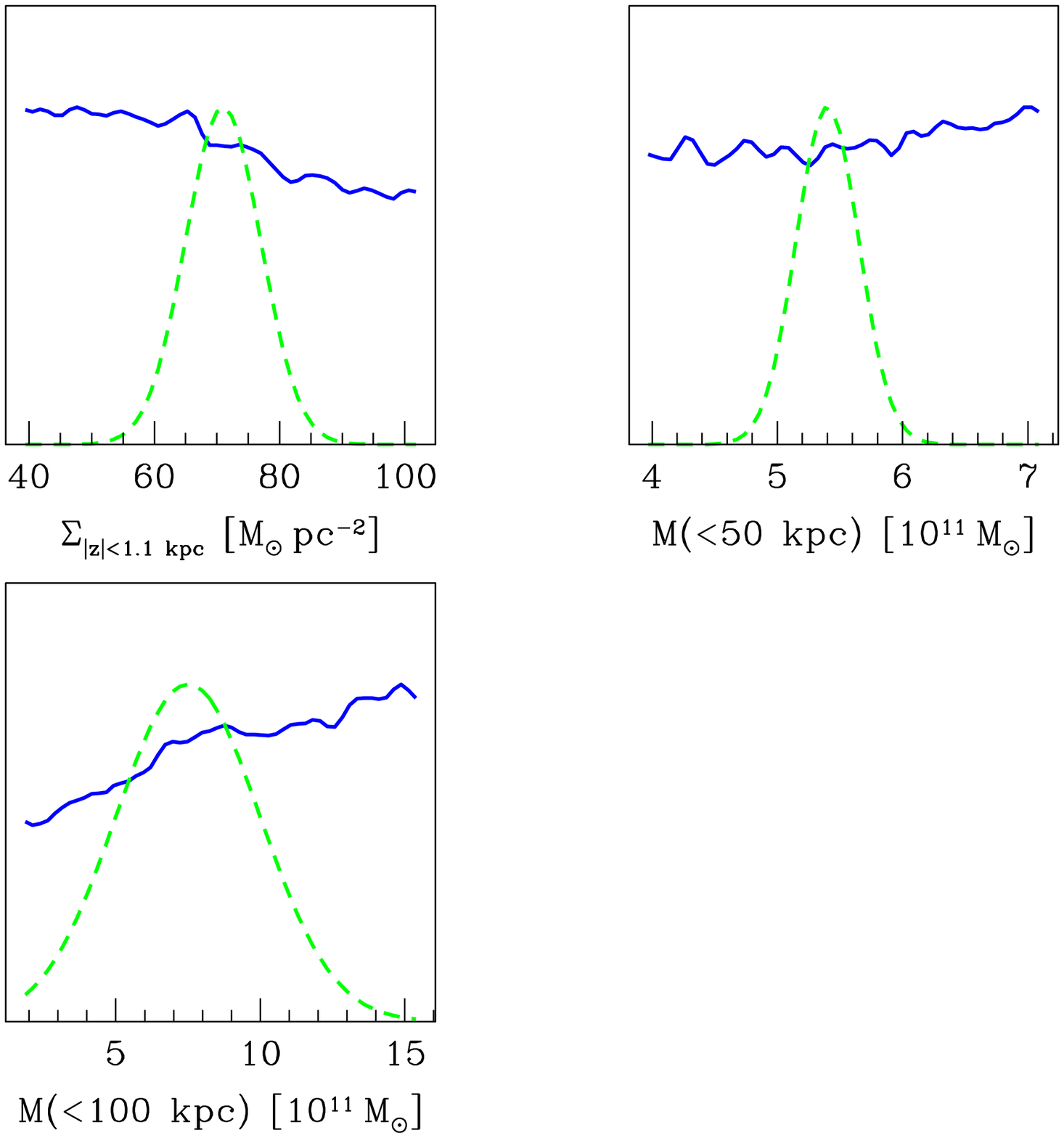}
\caption{\label{prior} The blue lines represent marginal posterior pdf for a few derived quantities obtained by a MCMC scan with no observational constraints. The green dashed curve are the corresponding Likelihoods. An Einasto profile has been assumed.}
\end{center}
\end{figure}

All the averages with respect to the different marginal pdf and the associated confidence intervals for the corresponding Galactic model parameters can be found in tables \ref{table_einasto} and \ref{table_nfw}. In all the one-dimensional plots discussed above, the red dashed lines represent the best fit value of the parameter considered in that plot, while the solid lines its average with respect to the plotted pdf. We say that a point in parameter space gives the best fit to the data when it minimizes the effective $\chi^2_{\rm eff}$ defined as follows
\beq
\chi_{\rm eff}^2= -2 \ln p(d|g) \,.
\eeq 
Means and best fit values differ considerably when the Likelihood surface is characterized by a narrow region in parameter space with a very high Likelihood surrounded by other regions with a slightly lower Likelihood but whose volume in parameter space is much larger.  In the present analysis, however, means and best fit values are rather close in all the plots, with the exception of the already mentioned bulge/bar central energy density. This suggests that our Likelihood contributes to the posterior pdf more then the prior distribution (see Eq.~(\ref{Bayes})). This indication is confirmed by Fig.~(\ref{prior}) which shows the marginal posterior pdf for a few derived quantities obtained through a MCMC scan performed without imposing any observational constraint. Such pdf represent the true priors which are actually used for the derived quantities. Since the derived quantities depend non trivially on the model parameters, the corresponding priors are not necessarily flat, even if flat priors in the parameter space are assumed \cite{Trotta:2008bp}. In our case, however, in the range where the Likelihoods are different from zero, it is always possible to disentangle the peaks of the Likelihoods from the priors on the derived quantities. Therefore, the observational constraints considered in our analysis are stringent enough to extract reliable informations from the data. 

\newpage
\section{Conclusions}
\label{concl}

We have produced a novel study on the problem of constructing mass models for the Milky
Way, concentrating on features regarding the dark matter halo component. We have
implemented
a variegated sample of dynamical observables for the Galaxy, including several results
which have
appeared recently. We have also developed our analysis introducing a rather general
scheme to
describe the different mass components for the Milky Way, introducing a model with a
large number
of parameters (8 or 7 in total). Compared to previous studies of this kind, which did
concentrate
on few sample choices of values for the different parameters, we have studied the full
parameter
space by implementing a Bayesian approach to the parameter estimation, analogously to
what is commonly done to estimate cosmological parameters, and a Markov Chain Monte Carlo
to
explore it. Our results show that this novel approach is fully successful.

The main result of this analysis is a novel determination of the local dark matter halo
density
which, assuming spherical symmetry and either an Einasto or an NFW density profile,
is found to be around 0.39~GeV~cm$^{-3}$ with a 1-$\sigma$ error bar of about 7\%; more
precisely we find a  $\rho_{DM}(R_0) = 0.385 \pm 0.027\,\rm GeV \,cm^{-3}$ for the
Einasto profile
and  $\rho_{DM}(R_0) = 0.389 \pm 0.025\,\rm GeV \,cm^{-3}$ for the NFW. This is in
contrast to
the standard assumption that $\rho_{DM}(R_0)$ is about 0.3~GeV~cm$^{-3}$ with
an uncertainty of a factor of 2 to 3. Our results indicate that this accuracy is
preserved even considering other spherical dark matter profiles such as the cored Burkert
profile for which we find  $\rho_{DM}(R_0) = 0.409 \pm 0.029\,\rm GeV \,cm^{-3}$. 
Sketching how solid this result is when introducing axisymmetric or triaxial
dark matter profiles
will be the goal of a dedicated study.

A very precise determination of the local halo density is very important for interpreting
direct dark matter detection experiments (see for instance the recent analysis \cite{trotta1}). Indeed the results we produced, together with the
recent
accurate determination of the circular velocity, should be very useful to considerably
narrow astrophysical uncertainties on direct dark matter detection.

\section*{Acknowledgment}
We would like to thank Joakim Edsjo for useful discussions during the early stages of the project, Gianfranco Bertone for interesting comments on our analysis and Roberto Trotta for suggesting the crosscheck associated to Fig.~(\ref{prior}). We are also grateful to Mark Reid to provide us a numerical package which transforms coordinates in the equatorial system to Galactic coordinates. The work of P.U. was partially supported by the EU FP6 "UniverseNet" Research \& Training Network under contract MRTN-CT-2006-035863.

\appendix

\section{Markov Chain Monte Carlo}

To calculate the integrals which define the means and the marginal pdf introduced in section \ref{Likelihood} is in principle a formidable task when the parameter space is of high dimensionality and the Likelihood surface exhibits a complicated structure. Monte Carlo methods based on a Makrkov Chain sampling are an usefull tool to overcome this complication. In this appendix we review this idea and show how it applies to the present analysis (see also {\it e.g.}, \cite{rewmcmc}). 

\subsection{Evaluation of expectation values and marginal pdf}

Let us introduce a generic $n$-dimensinal parameter space which can be though of as the parameter space of the Galactic models investigated in this paper. In the NFW case $n=7$, while in the case on an Einasto profile $n=8$. Let us now denote a point in such a parameter space by $g=(g_1,\dots, g_n)$. Given a sample $g^{(t)}$ of dimension N ($t=0,\dots,N-1$) drawn from an opportune posterior pdf, {\it e. g.} $p(g|d)$, a Monte Carlo estimate of an expectation value such as the one in Eq.~(\ref{ev}) is given by 
\beq
\langle f(g) \rangle = \int dg \,f(g)p(g|d)
\approx \frac{1}{N}\sum_{t=0}^{N-1} f(g^{(t)})\,\,.
\label{montecarlo}
\eeq
An analogous expression with $f(g)=1$ and the integration voulume reduced to an opportune subspace leads to the desired marginal pdf.

Thus, Monte Carlo methods reduce complicated integrals to much easier sums. The price to pay is the need of a smapling method. Sampling methods to obtain the sequence $g^{(t)}$ are typically based on the generation of Markov chains. Formally a Markov Chain is a sequence of random variables such that \cite{rewmcmc}
\beq
p(g^{(m+1)}|g^{(m)},\{ g^{(t)} : t \in \mathcal{S}\}) = p(g^{(m+1)}|g^{(m)}) \,,
\eeq   
where $\mathcal{S}$ is any subset of $\{ 0,\dots,m-1  \}$.
In other words, the first $m-1$ points of the chain only indirectly influence the probability of $g^{(m+1)}$. In practice a Markov chain is assigned giving the probability $p(g_{\rm in})$ that a point $g_{\rm in}$ is the starting point of the chain, and the so called transition probability $T_{m}(g,g^{\prime})$, which is the probability to find $g^{\prime}$ at the position $m+1$ in the chain, starting from $g$ at the position $m$. With these definitions, the probability of finding $g^{*}$ at the position $m+1$ is also given by $p_{m+1}(g^{*})= \sum_{\bar{g}} p_{m}(\bar{g}) T_{m}(\bar{g},g^{*})$.

One of the nice feature of Markov chains is that they admit invariant distributions, that is, distributions such that once they are realized at the position $m^{*}$ in the chain, they are not modified by further applications of the transition probability in all the subsequent positions $m>m^{*}$.  In general, an invariant distribution $I(g)$ is defined as follows
\beq   
I(g) = \sum_{\bar{g}}I(\bar{g}) T_{m}(\bar{g},g)  \qquad \forall \, m\,.
\eeq
A sufficient condition to ensure that a given pdf $\pi(g)$ is an invariant distributions of a Markov chain generated by a given transition probability is the condition of detailed balance, namely
\beq
\pi(g)T_{m}(g,g^{\prime}) = \pi(g^{\prime}) T_{m}(g^{\prime},g) \qquad \forall \, m\,.
\eeq

Once a Markov chain is correctly generated one can use it to estimate various expectation values by means of Eq.~(\ref{montecarlo}). Clearly, in the present analysis we are interested in generating Markov chains which have the posterior pdf (\ref{Bayes}) as invariant distribution. 

Our FORTRAN code uses the Markov Chain Monte Carlo routines of Superbayes \cite{superbayes} which generate Markov chains by means of the Metropolis algorithm. Given a starting point $g_{\rm in}$ drawn from an assigned distribution $p_{\rm in}(g_{\rm in})$, the Metropolis algorithm returns the subsequent point $g_{\rm out}$ proceding as follows. First, it proposes a new point $\bar{g}$ which is drawn from a given distribution $S(g_{\rm in},\bar{g})$, called the proposal distribution. Then, the proposed point $\bar{g}$ is accepted with probability
\beq
A(g_{\rm in},\bar{g}) = \min\left(1,\frac{p(\bar{g})}{p(g_{\rm in})}\right) \,. 
\eeq
If the proposed point $\bar{g}$ is accepted, the algorithm sets $g_{\rm out}= \bar{g}$. Otherwise, if rejected, the algorithm imposes $g_{\rm out} = g_{\rm in}$ and proposes a new point.  The number of proposed points $\bar{g}$ needed to move from $g_{\rm in}$ to $g_{\rm out}$ is the multiplicity assigned to $g_{\rm in}$. The sum (\ref{montecarlo}) counts each element of the chain, say $g^{(t)}$, with its multiplicity. It can be proven that after a number of iterations large enough, the Metropolis algorithm starts sampling from the desired pdf.

Concerning our choice of the proposal distribution, we followed the prescription of \cite{cosmomc}. 

\subsection{Convegence of Markov chains}

A set of Markov chains can be used to infer predictions only if they have converged to the desired invariant distribution, which for the present discussion corresponds to Eq.(\ref{Bayes}). We outline here the convergence criterion which we adopted in our analysis, refering to \cite{Gelman} for a more detailed treatement of the subject. 
 
For each halo type considered in this paper, we generated 24 chains containing 50000 points of the corresponding parameter space. Operationally, such chains have converged when inferences separately derived from any single chain lead to the same conclusions.  Given $l$ chains with $q$ elements sampled assuming a given dark matter profile, this idea can be formalized as follows. Let us focus on a generic model parameter $g_k$. We denote by $g^{ij}_k$ the value associated to $g_k$ after $j$ iterations within the $i$-th chain, where $i=1,\dots,l$ and $j=1,\dots,q$. The within-chains variance $W$ and between-chains variance $B/q$ of $g_k$ are defined by the relations  
\beqra
&&B/q = \frac{1}{l-1}\sum_{i=1}^{l}(\bar{g}_k^{i.}-\bar{g}_k^{..})^2 \nonumber\\
&&W = \frac{1}{l(q-1)}\sum_{i=1}^{l}\sum_{j=1}^{q}(\bar{g}_k^{ij}-\bar{g}_k^{i.})^2 \nonumber\\
\eeqra
where $\bar{g}_k^{i.} = \frac{1}{q}\sum_{j=1}^{q} g^{ij}_k$ and  $\bar{g}_k^{..} = \frac{1}{lq}\sum_{i=1}^{l}\sum_{j=1}^{q} g^{ij}_k$.

\begin{figure}
\begin{center}
\includegraphics[width=7.5 cm, angle=-90]{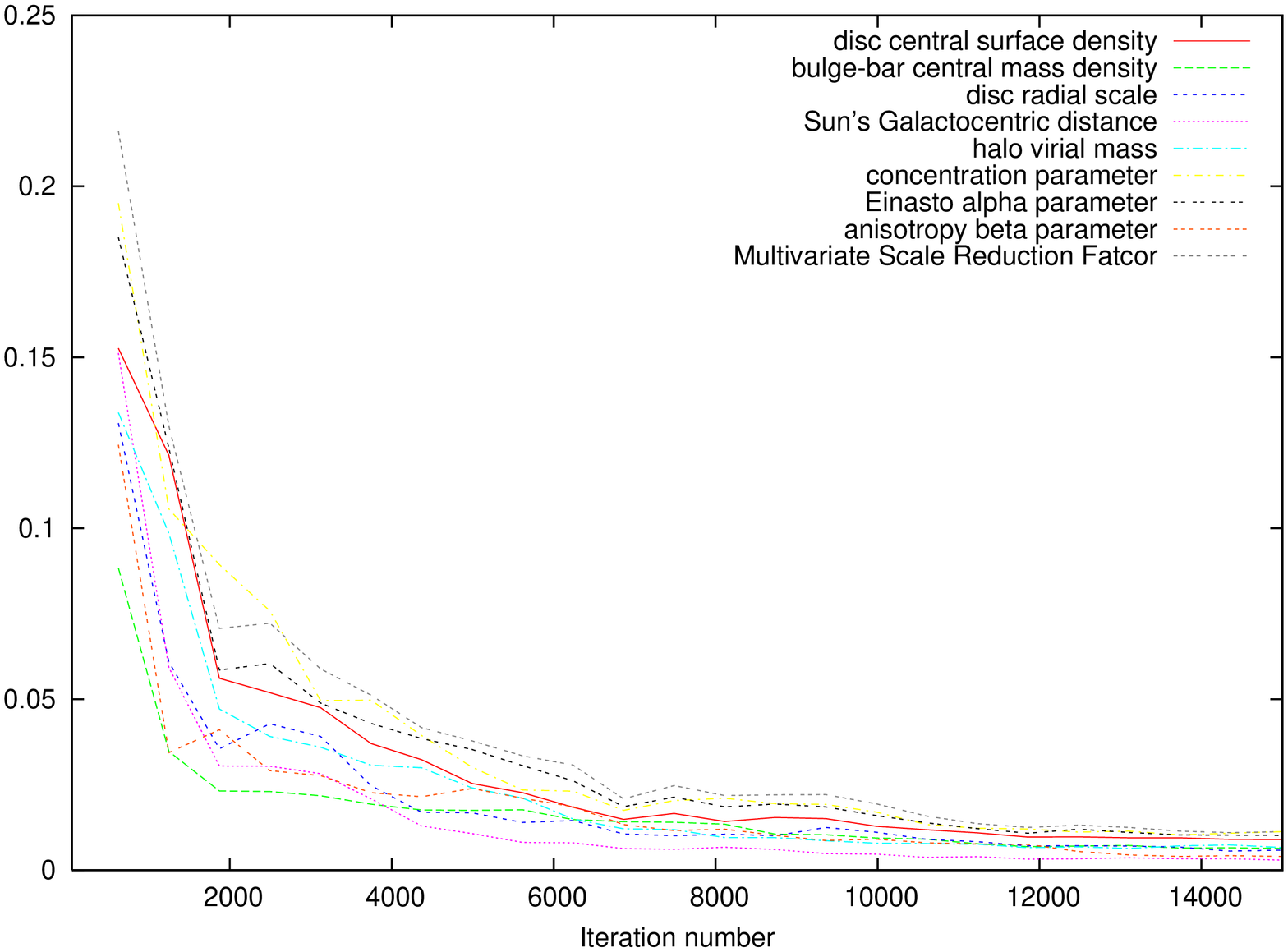}
\includegraphics[width=7.5 cm,angle=-90]{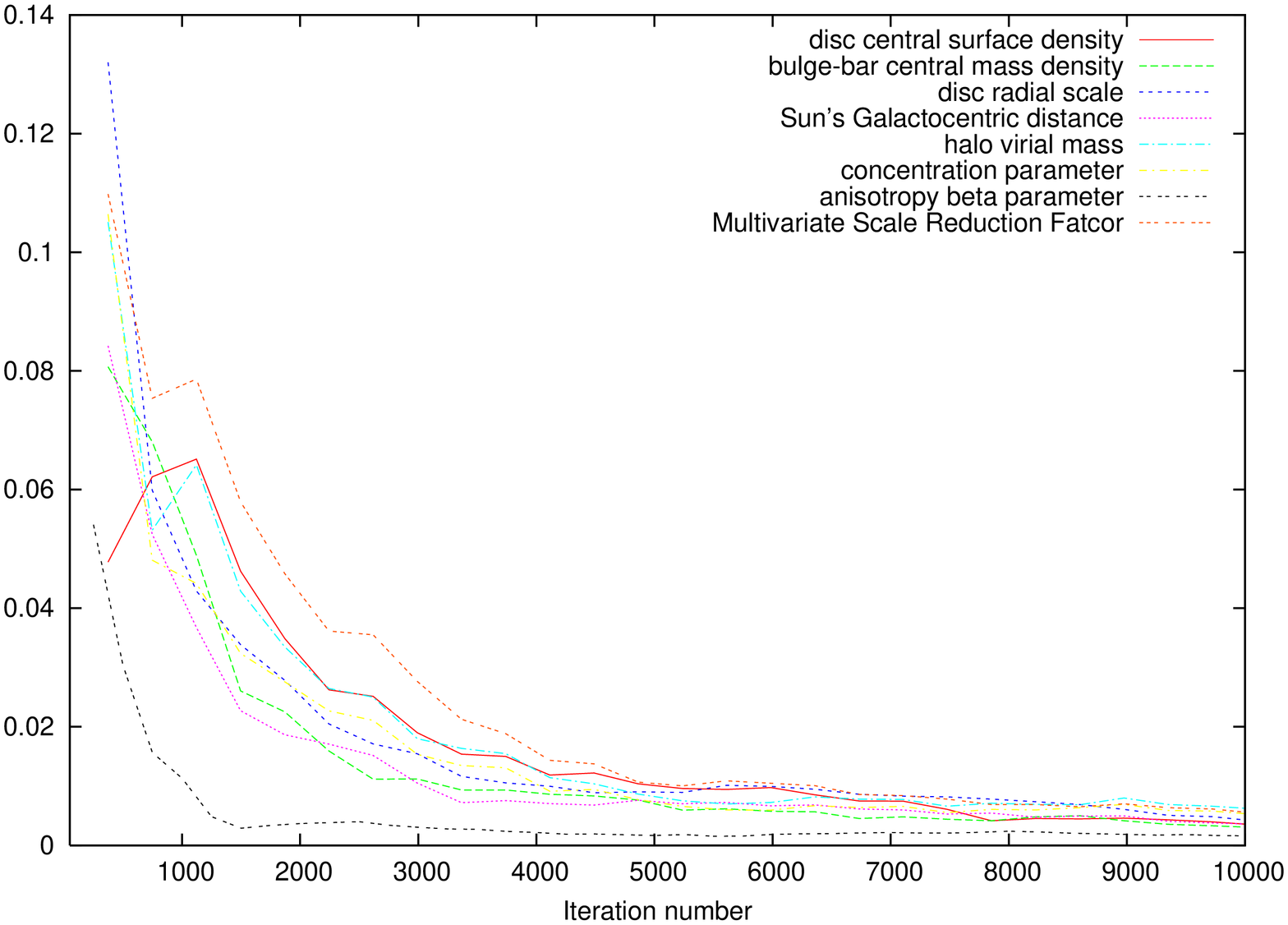}
\caption{\label{burnin} Top panel: Assuming an Einasto profile, we plot Eq.(\ref{ratio}) versus the iteraton number in the Metropolis algorithm for each parameter of the model. We also plot the  multivariate scale reduction factor \cite{Gelman2}. Bottom panel: As in the top panel but for the NFW case.}
\end{center}
\end{figure}

The weighted average of these quantities 
\beq
\sigma^{2}_{+}= \frac{q-1}{q} W + \frac{B}{q}
\label{WB}
\eeq
provides an estimate of the true variance $\sigma_{g_k}^{2}$ of the parameter $g_k$.  Indeed, if all the chains start from points well separed in parameter space, Eq.~(\ref{WB}) overestimates $\sigma_{g_k}^{2}$. On the other hand, the within-chains variance $W$ alone would underestimate $\sigma_{g_k}^{2}$. This argument suggests to monitor the ratio
\beq
R_k = \frac{\sigma_{+}^{2}+B/lq}{W} \,,
\label{ratio}
\eeq
considererd as a function of the iteration number of the Metropolis algorithm. The term $B/lq$ in the numerator is a correction to the between-chains variance which accounts for the sampling variability \cite{Gelman2}. By construction, the ratio (\ref{ratio}) is initially larger then one. However, when the chains have converged and the Metropolis algorithm starts sampling from the desired posterior pdf, $R_k$ stabilizes around a constant value close to one. This is the case of the Markov chains generated in our analysis (see Fig. \ref{burnin}). We also considered a convergence criterion based on the so called multivariate scale reduction factor $R$. This approach extends the present method, which applies to each parameter $g_k$ separately, to the case in which all the parameters 
are treated together. We refer to \cite{Gelman2} for more details on the multivariate approach.           

\clearpage
\begin{table}[c]
  \centering
  \renewcommand{\arraystretch}{1.3}
  \begin{tabular}{|c|c|c|c|c|}
    \hline
    Galactic components & Parameters & Units & $g_k^{\rm min}$ & $g_k^{\rm max}$\\
    \hline \hline
    Disc & $\Sigma_{d}$ & $M_{\odot}\,pc^{-2}$ & 400 & 2000 \\ 
    Disc & $R_d$ & kpc & 1.2 & 4 \\ 
    Dust layer & none & - & - & -\\ 
    Bulge/bar & $\rho_{bb}(0)$ & $M_{\odot}\,pc^{-3}$ & 0.1 & 3 \\
    Halo & $M_{vir}$ & $10^{12} M_{\odot}$ & 0.1 & 10 \\
    Halo & $c_{vir}$ & - & 5 & 25 \\
    Einasto halo & $\alpha_E$ & - & 0.1 & 0.4 \\
    All components & $\beta$ & - & -1 & 0.7 \\
    All components & $R_0$ & kpc & 6.5 & 9.5 \\  
    \hline 
  \end{tabular}
  \caption{Parameters of the model. See the text for their physical interpretation.
           \label{priors}}
 \end{table}


\clearpage

\begin{table}[t]
  \footnotesize
  \centering
  \renewcommand{\arraystretch}{1.}
  \begin{tabular}{|c|c|c|c|c|c|c|}
    \hline
   Parameters               &     mean      &  $\sigma$ & low 68.00\% & up  68.00\% & low 95.00 \% & up 95.00 \% \\
    \hline \hline
 $\Sigma_d$ [M$_{\odot}$pc$^{-2}$~]    &   1154.14 &   427.43 &  683.14 &  1662.94 &  476.17 &  1943.67 \\  
 $\rho_{bb}(0)$ [M$_{\odot}$pc$^{-3}$~]&   1.37    &  0.80    & 0.48    &  2.32    &  0.16   &  2.9 \\ 
 $R_d$ [kpc]                          &   2.45    &  0.21    & 2.24    &  2.65    &  2.07   &  2.90 \\ 
 $R_0$ [kpc]                          &   8.25    &  0.29    & 7.97    &  8.54    &  7.66   &  8.81  \\ 
 $M_{vir}$ [10$^{12}$~M$_{\odot}$]      &   1.39     &  0.33   & 1.07    &  1.74    & 0.93     & 2.14   \\ 
 $c_{vir}$                             &   18.01    &  3.32   & 14.51   &  21.76   & 12.31   & 24.36   \\ 
 $\alpha_E$                            &   0.22    &  0.07   & 0.14     & 0.29     & 0.11    & 0.35    \\ 
 $\beta$                               &  -0.29    &  0.24   &-0.53     &-0.06     &-0.80    & 0.13    \\ 
  \hline \hline
  Derived quantities        &     mean   &  $\sigma$ & low 68.00\% & up  68.00\% & low 95.00 \% & up 95.00 \% \\
  \hline \hline
 $A-B$  [km~s$^{-1}$~kpc$^{-1}$]               &   29.44   &  0.15   & 29.29    & 29.59    & 29.15   & 29.74   \\ 
 $A+B$  [km~s$^{-1}$~kpc$^{-1}$]               &   0.07    &  0.45   &-0.38     & 0.51     &-0.82    & 0.94    \\ 
 $v_c(R_0)$ [km~s$^{-1}$]                      &   243.03  &  8.49   & 234.68   & 251.28   & 225.61  & 259.13  \\ 
 $\Sigma_*$ [M$_{\odot}$~pc$^{-2}$]             &   46.51   &  5.47   & 41.05    & 51.96    & 35.76   & 57.23   \\ 
 $\Sigma_{|z|<1.1\rm kpc}$ [M$_{\odot}$~pc$^{-2}$] &   72.16   &  4.24   & 67.93    & 76.37    & 63.87   & 80.47   \\ 
 $M(<50 \rm kpc)$ [10$^{11}$~M$_{\odot}$]       &  5.36    &  0.24   & 5.13     & 5.60     & 4.90    & 5.83    \\ 
 $M(<100 \rm kpc)$ [10$^{11}$~M$_{\odot}$]      &  8.59    &  0.64   & 7.94     & 9.23     & 7.37    & 9.86    \\ 
 $\rho_{DM}(R_0)$  [ GeV~cm$^{-3}$]             &  0.386    &  0.027  & 0.359    & 0.413    &  0.333  &  0.439 \\ 
    \hline 
  \end{tabular}
  \caption{Means, standard deviations and confidence intervals for the model parameters and the derived quantities in the Einasto case. \label{table_einasto}}
\end{table}

\begin{table}[c]
  \centering
  \footnotesize
  \renewcommand{\arraystretch}{1.}
  \begin{tabular}{|c|c|c|c|c|c|c|}
    \hline
   Parameters               &     mean   &  $\sigma$ & low 68.00\% & up  68.00\% & low 95.00 \% & up 95.00 \% \\
    \hline \hline
  $\Sigma_d$ [M$_{\odot}$pc$^{-2}$~]      &   1042.62 &   188.91  &  849.17       &  1231.13      & 678.60         & 1404.01  \\ 
  $\rho_{bb}(0)$ [M$_{\odot}$pc$^{-3}$~]  &   1.31    &   0.79    & 0.44          &  2.25         & 0.15           & 2.85     \\ 
  $R_d$ [kpc]                            &   2.41    &  0.17     & 2.23          &  2.58         & 2.08           & 2.75     \\ 
  $R_0$ [kpc]                            &   8.28    &  0.29     & 8.00          &  8.55         & 7.67           & 8.81     \\ 
  $M_{vir}$ [10$^{12}$~M$_{\odot}$]        &    1.49   &  0.17     & 1.33          &  1.64         & 1.23            & 1.86     \\ 
  $c_{vir}$                              &    19.70   & 2.92     & 16.59         &  22.90        & 13.93           & 24.60    \\ 
  $\beta$                                &  -0.30    &  0.23     &-0.53          &-0.072         &-0.79            & 0.11     \\ 
   \hline \hline
   Derived quantities               &     mean   &  $\sigma$ & low 68.00\% & up  68.00\% & low 95.00 \% & up 95.00 \% \\
    \hline \hline
  $A-B$ [km~s$^{-1}$~kpc$^{-1}$]         &   29.44   &  0.15     & 0.29.29       & 0.29.59       & 29.15           & 29.74    \\ 
  $A+B$ [km~s$^{-1}$~kpc$^{-1}$]         &   0.073   &  0.44     &-0.37          & 0.51          &-0.79            & 0.94     \\ 
  $v_c(R_0)$  [km~s$^{-1}$]              &    243.75  &  8.34     & 235.58        & 251.79        & 226.11          & 259.05   \\ 
  $\Sigma_*$  [M$_{\odot}$~pc$^{-2}$]     &   46.24   &  5.38     & 40.87         & 51.60         & 35.77           & 56.87    \\ 
  $\Sigma_{|z|<1.1 \rm kpc}$ [M$_{\odot}$~pc$^{-2}$] &    72.13   &  4.18     & 67.95         & 76.29         & 63.93   & 80.31    \\ 
  $M(<50 \rm kpc)$ [10$^{11}$~M$_{\odot}$]  &   5.35    &  0.24     & 5.11          & 5.59          & 4.88            & 5.82     \\ 
  $M(<100 \rm kpc)$ [10$^{11}$~M$_{\odot}$] &   8.56    &  0.53     & 8.035         & 9.08          & 7.59            & 9.65     \\ 
  $\rho_{DM}(R_0)$ [GeV~cm$^{-3}$]         &   0.389   &  0.025    & 0.365         & 0.414         & 0.338           & 0.435     \\ 
    \hline 
  \end{tabular}
  \caption{Means, standard deviations and confidence intervals for the model parameters and the derived quantities in the NFW case.
 \label{table_nfw}}
\end{table}

\end{document}